\documentclass[12pt,preprint]{aastex}

\usepackage{amsmath}

\begin{document}





\title{On the formation of Be stars through binary interaction
}
\author{
Yong Shao$^{1}$ and Xiang-Dong Li$^{1,2}$}

\affil{$^{1}$Department of Astronomy, Nanjing University, Nanjing 210093, China}

\affil{$^{2}$Key laboratory of Modern Astronomy and Astrophysics (Nanjing University), Ministry of
Education, Nanjing 210093, China}

\affil{$^{}$lixd@nju.edu.cn}

\begin{abstract}

Be stars are rapidly rotating B type stars. The origin of their rapid
rotation is not certain, but binary interaction remains to be a possibility.
In this work we investigate the formation of Be stars
resulting from mass transfer in binaries in the Galaxy.
We calculate  the binary evolution with both stars evolving
simultaneously and consider different possible mass accretion
histories for the accretor. From the calculated results we obtain
the critical mass ratios  $q_{\rm cr}$ that determine the stability
of mass transfer. We also numerically calculate the parameter
$\lambda$ in common envelope evolution, and then incorporate both
$q_{\rm cr}$ and $\lambda$ into the population synthesis
calculations. We present the predicted numbers and characteristics
of Be stars in binary systems  with different types of companions,
including helium stars, white dwarfs, neutron stars, and black
holes. We find that in Be/neutron star binaries the Be stars can
have a lower limit of mass $ \sim 8 M_{\odot}$ if they are formed by
stable  (i.e., without the occurrence of common envelope evolution)
and nonconservative mass transfer. We demonstrate that isolated Be
stars may originate from both mergers of two main-sequence stars and
disrupted Be binaries during the supernova explosions of the primary
stars, but mergers seem to play a much more important role. Finally
the fraction of Be stars which have involved binary interactions in
all B type stars can be as high as $ \sim 13\%-30\% $, implying that
most of Be stars may result from binary interaction.

\end{abstract}


\section{Introduction}

Be stars are rapidly rotating B type stars with luminosity classes
III-V, which show $\rm H\alpha$ emission lines and excess infrared
fluxes in some intervals of their lives. The characteristics of Be
stars is thought to originate in the circumstellar disk
\citep{pr03}. Their rotational velocities generally reach nearly
break-up velocities, leading to the formation of a gaseous decretion
viscous disk around them
\citep[e.g.,][]{l1991,w1997,o2001,pr03,c2009,j2009,s2009,m2013}. The
Be stars have been found either as single stars or in binary
systems. In Be/X-ray binaries (BeXRBs), a compact star, usually a
neutron star (NS), orbits a Be star and accretes the dense stellar
wind from the Be star, therefore emitting X-rays. In the Galaxy
there are 81 BeXRBs, and 48 out of them have been found to host a NS
\citep{l06,reig11}. They have been detected as X-ray sources with
luminosities in the range of $ \sim 10^{34}-10^{38}\, \rm erg$ $\rm
s^{-1}$, and the orbital periods range from $ \sim10 $ days to
several hundred days \citep{reig11}. Recently \citet{cn14}
discovered a Be/black hole (BH) binary MWC 656 with an orbital
period $\sim 60$ days, but its X-ray emission is extremely weak
\citep{ma14}. It is interesting to note that the Be stars in BeXRBs
have  spectral types earlier  than B2 \citep{n98}, while isolated Be
stars have a spectral distribution of A0$ - $O9 \citep{s82}.

The formation of Be stars is still a controversial topic.
There are three possible explanations for the origin of the rapid
rotation of Be stars \citep[e.g.,][]{h10}: (1) they were born as
rapid rotators; (2) along the main-sequence (MS) evolution, single
stars with a sufficiently high rotational velocity on the zero-age
main-sequence (ZAMS) can reach equatorial velocities near the
critical value, due to the transfer of angular momentum from the
inner contracting part to the outer region \citep{e08}; and (3) they
are the mass gainers that were spun up by a past episode of
Roche-lobe overflow (RLOF) in an interacting binary
\citep{rh82,p91}. During the process of RLOF, matter and angular
momentum are transferred from the primary star to the secondary
star, spinning up the latter to very high rotation rates
\citep{p81}, which then turns into a Be star. According to
\citet{mg05}, about 75\% of the detected Be stars may have been
spun-up by binary mass transfer, while most of the remaining Be
stars were likely rapid rotators at birth. \citet{h10} showed that
most young B type stars have rotational velocities that are well
below the limit for Be star formation, suggesting that only a small
fraction of Be stars were born as rapid rotators.

Using a binary population synthesis (BPS) method, \citet{p91}
investigated the formation of Be stars by case B mass
transfer\footnote{The case of the mass transfer is a classification
of the mass transfer by the evolutionary status of the donor: Case A
- core hydrogen burning, Case B - shell hydrogen burning, Case C -
after exhaustion of core helium burning.}, and presented a
comparison between the predicted number of Be stars and
observations. They concluded that no more than 60\% of the
population of Be stars can be produced by close binary interaction.
\citet{p91} also attempted to address the observed lower mass limit
of $ \sim 8M_{\odot} $ for the Be stars in BeXRBs, corresponding to
the spectral type B2. They suggested that only systems with initial
mass ratios (i.e., the ratio of the secondary mass and the primary
mass) larger than $0.3-0.5 $ can produce Be stars. The argument was
that systems with smaller initial mass ratios do not transfer any
mass stably but evolve into a spiral-in common envelope (CE) phase,
without forming Be stars. Thus Be stars with spectral type later
than B2 can be effectively removed. In the calculations, they
adopted a simple assumption that the supernova (SN) explosion of the
primary is spherically symmetric, which helps the survival of the
binary after the SN.

\citet{pz95} considered the effect of an asymmetric SN explosion and used a
constant value for the kick velocity. Although the number of late-type Be stars
with a NS companion is reduced slightly, the spectral distribution of the Be stars
was found not to match the observations. The author then introduced possible mass loss from the
binary system at the second Lagrangian point $ L_{2} $ during the mass transfer
processes, with the escaped matter taking away $ \sim 6 $ times
the specific angular momentum of the binary system.
Because of the high rate of  angular momentum loss, the systems
with small initial mass ratios would also undergo spiral-in
and evolve towards a CE phase. Therefore Be stars with a NS companion should be
more massive than $ \sim 8 M_{\odot} $. \citet{bv97} also made population
synthesis calculation on the formation of Be stars via binary evolution
in the Galaxy and the Magellanic Clouds, incorporating updated data of the
SN kick. They assumed that a minimal initial mass ratio $ q_{\rm min} =0.2$
for stable mass transfer, below which CE evolution will occur.
Consequently the Be stars evovled from close binary evolution were shown to
contribute not too much to the total population of Be stars (less than 20\% and possibly even
as low as 5\%). A more recent BPS study of Galactic
BeXRBs was performed by \citet{b09}, who attempted to explain the
problem of the missing BeXRBs with BHs at that time. However, the discovery
of MWC 656 suggests that the population of Be/BH binaries are
not negligible in the Galaxy.

As mentioned above, the critical mass ratio $q_{\rm cr}$, which is
used to determine whether a binary system experiences stable mass
transfer, is one of the vital factors in the formation of the Be
stars through binary interaction. It depends closely on the
structure of the donor, the nature and the mass of the accretor, and
how conservative the mass transfer is
\citep[e.g.,][]{p92,prp02,kw96,spv97,lv97,t00,han02,it04,ge10,w12,sl12}.
Constant values or empirical formulae for $q_{\rm cr}$ are usually
adopted in the BPS calculations. In this work we numerically
calculate the critical mass ratios for various initial conditions,
and use them as input in the simulations of binary evolutions that
form Be stars.
\citet{d13} recently investigated the evolution of the rotation
rates of massive stars, and concluded that mass transfer and mergers
are the main cause of rapid rotation for massive stars. Here we
focus on the formation of Be stars, following some of their
treatments on binary interaction, especially the processes of
mergers. We introduce the calculating method and the input
parameters in Section 2. In Section 3 we present the calculated
results on the properties of Be stars both in binaries and as single
stars, and compare them with observations. We discuss the fraction
of Be stars in B type stars contributed by binary evolution and
conclude in Section 4.


\section{Method}

\subsection{The binary population synthesis code}

We adopt the BSE code initially developed by \citet{h00,h02}
and revised by \citet{kh06} to calculate the evolution of a large population of massive stars.
We follow \citet{b08} to update some of the treatments
of the processes that lead to
the formation and evolution of compact objects. In addition,
we have modified the code in the following aspects.

We adopt the rapid supernova (SN) mechanism \citep{w02,f12} to obtain the mass
of a NS/BH after a SN explosion, which seems
to account for the combined mass distribution of NSs and BHs,
with a dearth of the remnants of mass between $\sim 2 M_{\odot}$ and
$\sim 4-5 M_{\odot}$ \citep{ozel10,farr11}.
The final mass of a compact object is determined
by the CO core mass at the time of explosion, which gives
the proto-compact object mass. In the subsequent explosion, accretion of the
fallback material increases its mass to form a NS or BH. For
electron-capture SNe, we apply the criterion suggested
by \citet{f12} as follows.
If the core mass at the base of asymptotic giant branch is between
$ 1.83 M_{\odot} $ and $ 2.25 M_{\odot} $, the CO core non-explosively
burns into an ONe core and the core mass increases gradually.
If the core mass can reach $1.38 M_{\odot} $, the core
collapses by electron capture into Mg and forms a NS. If the
ONe core mass is less than $1.38 M_{\odot}$, it leaves an ONe WD.

The natal kick imparted on the newborn compact objects is an important factor
that determines the formation efficiencies of XRBs.
We adopt  a Maxwellian distribution for the kick velocity with
$ \sigma = 265\, \rm km\, {\rm s}^{-1} $ \citep{h05}
for NSs formed from core-collapse SNe.
For electron-capture SNe, we take a lower kick velocity
with $ \sigma = 50\, \rm km\, {\rm s}^{-1} $ \citep{d06}. For BHs, if the fallback
material fraction $ f_{\rm fb}=1$ (i.e., direct collapse), there is no natal kick.
Otherwise we use the NS kick velocity reduced by a factor of $ (1- f_{\rm fb}) $
for $ f_{\rm fb} < 1$ \citep{f12}.

If the mass transfer is dynamically unstable during the ROLF, a binary
will enter the CE phase.
We use the standard energy conservation equation
\citep{w84} to deal with the CE evolution,
\begin{equation}
\alpha_{\rm CE}(\frac{GM_{\rm 1,f}M_{2}}{2a_{\rm f}}
-\frac{GM_{\rm 1,i}M_{2}}{2a_{\rm i}}) =-E_{\rm bind},
\end{equation}
and
\begin{equation}
E_{\rm bind} = -\frac{GM_{\rm 1,i}M_{\rm 1,env}}{\lambda R_{\rm 1,lobe}},
\end{equation}
where $M_1$ and $M_2$ are the primary (mass donor) and secondary
(mass gainer) masses respectively, $a$ is the orbital separation of
the binary, $ M_{\rm 1,env} $ is the mass of the primary's envelope
that is ejected from the system during the CE evolution, $ R_{\rm
1,lobe} $ is the RL radius of the primary at the onset of RLOF, and
the indices i and f refer to the initial and final stages of the CE
evolution, respectively. The parameter $ \lambda $ includes the
effect of the mass distribution within the envelope and the
contribution from the internal energy \citep{kool90,dewi00}, and $
\alpha_{\rm CE} $ is the CE efficiency with which the orbital energy
is used to unbind the stellar envelope. We employ the results in
\citet{xl10} to calculate $\lambda$, and take $ \alpha_{\rm CE} $ =
1.0 in our calculations.

\subsection{The critical mass ratio}

The critical mass ratio $ q_{\rm cr} $  can be used to
determine whether the mass transfer is dynamically unstable in a binary.
Instead of using the empirical results in \citet{h02} we numerically calculate
the values of $q_{\rm cr}$ and incorporate them into the BPS code.
Note that in this work we define the mass ratio $q=M_1/M_2$, different from
in previous studies.

We use an updated version of the stellar evolution code developed by
\citet{e71,e72} \citep[see also][]{p95,ye05} to calculate the binary
evolution, and search the parameter space for stable mass transfer.
In these calculations, we adopt the TWIN mode in which the structure
and the composition equations for both stars, as well as the orbital
properties such as the eccentricity and orbital angular momentum,
are solved simultaneously. The initial chemical compositions are set
to be $X = 0.7$ and $ Z=0.02 $. We take the ratio of the mixing
length to the pressure scale height to be 2.0, and the convective
overshooting parameter to be 0.12 \citep{s97}. For the wind mass
loss rates from massive stars, we take the empirical formula for
luminous stars suggested by \citet{dnv88}.

\subsubsection{The models of mass transfer}

All of the binary stars considered here initially consist of two
ZAMS stars. The effective radius $ R_{\rm L,1} $ of the primary's RL
is given by \citep{e83}
\begin{equation}
\frac{R_{\rm L,1}}{a} = \frac{0.49q^{2/3}}{0.6q^{2/3} + \ln(1 + q^{1/3})}.
\end{equation}
We assume that the initial binary orbit is
circular \citep{k88}, and the orbital angular momentum is
\begin{equation}
J_{\rm orb} = \frac{M_{1}M_{2}}{M_{\rm T}}\omega a^{2},
\end{equation}
where $M_{\rm T} =M_{1}+M_{2}$ is the total mass, and $\omega=\sqrt{GM/a^{3}}$ is
the orbital angular velocity.

In order to follow the evolution of the orbit and spins of the two
stars in a binary, we consider the orbital angular momentum and the
spin angular momentum of each binary component. We assume that the
rotation of the stars is rigid. The coupling of the orbit and the
spins is controlled by tidal interaction, and we account for the
spin-orbit interaction by using the equilibrium tide theory
\citep{h81}. The code includes the loss of angular momentum by
stellar winds and the transfer of angular momentum between the two
stars.

As RLOF occurs, mass transfer onto the secondary will cause it to
expand and spin up. The secondary is also rejuvenated due to
accretion \citep{h02}. Several authors
\citep[e.g.,][]{ub76,n77,pm94} have investigated the evolution of
accreting MS stars and obtained the following results. If the mass
transfer time scale $\tau_{\dot{M}} (=-M_1/\dot{M}_1$, where
$-\dot{M}_1$ is the mass transfer rate) is longer than the thermal
time scale $\tau_{KH2}$ of the accretor, the mass transfer is
stable, and the mass gainer can remain in thermal equilibrium. On
the other hand, if $\tau_{\dot{M}}<\tau_{KH2} $, the accretor will
get out of thermal equilibrium and expand. This expansion may
finally cause the accretor to fill its own RL, leading to the
formation of a contact binary \citep[e.g.,][]{ne01}. The conditions
for thermal-timescale RLOF imply that the donor is more massive than
the accretor and possesses a radiative envelope, i.e., mass transfer
occurs in case A or early case B phases. Dynamically unstable mass
exchange usually occurs when the donor has developed a deep
convective envelope, i.e., mass exchange in late case B or case C
phases.

\citet{p81} pointed out
that only a small amount of accreted mass can spin up the accretor
to critical rotation. It is still unclear
whether and how a rapidly rotating star can keep accreting mass.
\citet{p05} and \citet{d09} assumed that mass accretion ceases when
the accretor reaches the critical rotation. Alternatively,
\citet{d13} suggested that a star can continue to accrete even with
the critical rotation, based on the argument of \citet{pa91} and
\citet{pn91} that the accretion disk can regulate the mass and
angular momentum flux through viscous coupling.

In summary, mass accretion can spin up the accretor, cause it to
expand, and probably result in mass loss. Since the process is rather complex,
here we construct three models to investigate the stability of mass transfer,
using the Eggleton's code to follow the response of the accretor.

\textit{Model I: rotation-dependent mass accretion}

We adopt the suggestion by \citet{se09} to deal with the accretion
of rapidly rotating stars. It is assumed that
the accretion rate onto a rotating star is reduced by a factor
of $ (1 - \Omega/\Omega_{\rm cr} $),
where $ \Omega $ is the angular velocity of the star and $ \Omega_{\rm cr} $
is its critical value. In this model a star rotating at
$ \Omega_{\rm cr} $ will not  accrete mass anymore, and we assume that the
remaining material is ejected out of the binary
in the form of isotropic wind, carrying the accretor$'$s
specific orbital angular momentum
$j_{\rm iso}=(M_{1}/M_2)(J_{\rm orb}/M_{\rm T})$.

\textit{Model II: half mass accretion and half mass loss}

We do not consider the detailed effects of rotation, and assume that
half of the transferred mass is accreted by
the secondary, and the other half is lost from the system, also taking the
specific orbital angular momentum of the accretor \citep[see also][]{d07}.

\textit{Model III: thermal equilibrium limited mass accretion}

In this case we assume that the transferred mass is always
accreted by the secondary unless its thermal timescale
becomes much shorter than the mass transfer timescale. Specifically, the
accretion rate is assumed to be limited by
$ -[\min (10\frac{\tau_{\dot{M}}}{\tau_{KH2}},1)]\dot{M}_1$ \citep{h02}.
Rapid mass accretion may drive the accretor out of thermal equilibrium,
which will expand and
become overluminous.  In our calculations, the values of $ \tau_{KH2} $
are found to be usually much lower than that of the
same star in thermal equilibrium, so that
$ \tau_{KH2} < 10\tau_{\dot{M}} $ always holds, meaning that mass
transfer is generally conservative.

Note that Models I and III represent two extreme cases of mass transfer,
corresponding to highly non-conservative [only a small fraction
($ \lesssim 10\% $) of the transferred material is accreted by the secondary]
and roughly conservative mass transfer, respectively. Model II describes an
intermediate between them, with $ 50\% $ of the transferred mass accreted.


In our calculations the binary evolution will be stopped when either
the radius of the accretor exceeds its RL radius or the mass
transfer rate rises  rapidly to very high rate
($>10^{-3}\,M_\sun$yr$^{-1}$) and the code fails to converge. The
binary is then assumed to become contact or enter the CE phase. The
evolution of contact binaries, however, is not yet fully understood.
It may be driven not only by mass transfer but also by luminosity
transfer between the components \citep{sl81}. We follow the previous
suggestion that the fate of contact binaries consisting of two MS
stars is a merger, forming a single star with rapid rotation
\citep{d07,d13,j13}. For the contact binaries containing a
Hertzsprung gap (HG) donor, \citet{d13} assumed that they will merge
to become blue or red supergiants. Here we assume that they will
evolve into the CE phase following \citet{h02}, and whether the
binary components will merge is determined by the energy equation in
Section 2.1.

\subsubsection{The parameter space for stable mass transfer}

We have calculated the mass transfer processes in a grid of binaries with different
values of the initial parameters. The primary masses $ M_{1} $ are taken
to be 1.5, 3, 5, 7, 10, 20, 30, 40, 50, and 60 $ M_{\odot} $.
The orbital periods $P_{\rm orb}$ (in units of days) vary
logarithmically from $-0.5$ to 3.5 by steps of 0.1. If the initial
orbital period is so short that the primary has filled its RL at the
beginning of binary evolution, it will skip to the next longer
orbital period. The mass ratios $q$ are increased from 1.2 to 6 by
steps of 0.2$-$0.5.

In Fig.~1 we outline the boundaries that determine whether a binary
can evolve successfully with stable mass transfer in the initial
$P_{\rm orb}-M_{1}$  plane. The left, middle and right panels
correspond to the results in Models I, II and III, respectively. The
values of initial $ q $ are indicated with different colors in the
figure. The solid curves, which always appear in pairs, show the
lower and upper boundaries of the parameter space for a specific
$q$, between which the mass transfer can proceed stably. We also
plot other curves to distinguish the evolutionary states of the
donor star - the thick grey curve (which overlaps with the curves
for the upper boundary in the left panel) denotes the orbital
periods when the primary fills its RL at the end of HG; the green
dashed and dotted curves represent the orbital periods when the
RL-filling primary is at the beginning and at the end of MS,
respectively.

We see from Fig.~1 that the mass transfer always becomes
runaway when the primary has climbed to the (super)giant branch and developed
a deep convective envelope prior to the mass exchange.
In Model I, expansion of the mass gainer is not significant
because of small amount of mass accreted. Loss of mass and angular momentum
from the binary shrinks the orbit before the mass ratio reverses.
For a binary with sufficiently short $ P_{\rm orb}$,
this may lead to the accretor$'$s radius exceeding its RL radius.
The value of $ q_{\rm cr} $ for stable mass transfer can reach as
high as $ \sim 6 $. In Model II, half of the transferred material is
assumed to leave the binary. The regions for stable mass transfer
seem to be odd, with isolated ``islands" in the parameter space in
some cases, implying that the stability of the mass transfer is
sensitive to the orbital periods and the masses of the binary
components. Generally $ q_{\rm cr} \lesssim 2.5$ in this case. In
Model III the secondary accretes more material from the primary than
in Models I and II, and expands more significantly, so the parameter
space for stable mass transfer without contact is smaller. We can
see that for a HG star with $M_1 \gtrsim 20 M_{\odot}$, a contact
phase always occurs. In this situation $ q_{\rm cr} \lesssim 2.2$.
Generally the lower the mass ratio, the bigger the allowed parameter space is.
In the appendix we present two examples of the evolutionary tracks,
to demonstrate how the mass transfer depends on the initial parameters
and mass loss modes.

\subsection{Possible channels to form a Be star}
Although Be stars are thought to be rapidly rotating B type stars,
it is controversial how fast  a B star can spin before it becomes a
Be star. To  calculate the number and distribution of Be stars in
the Galaxy, we adopt a phenomenological definition of Be stars based
on their observational spectral types and characteristics, i.e.,
they are MS stars of mass between $3\,M_{\odot} $ and
$22\,M_{\odot} $, rotating at $ \geq 80\% $ of their break-up
velocities $V_{\rm cr}$ \citep{s82,n98,pr03}.

Here we consider the binary interaction
to form Be stars involving stellar winds, tides, mass transfer and
mergers \citep[see also][]{d13}. We follow the treatments on these
processes in \citet{h02} and \citet{kh06} to calculate the stellar
rotation in binary systems.

\textit{Stellar evolution}. When there is angular momentum transfer
between the inner and outer parts of a star, the stellar rotation is
decided by the stellar structure. The moment of inertia of the star,
$ I = kMR^{2}$, where $ k $ is the radius of gyration squared, is
given by Pols in form of fitting formulae
\citep[see][]{d13}, and added in the BSE code. For ZAMS
stars, the gyration radius squared $ k_{0} $ is given by
\begin{equation}
k_{0} \simeq  c +  \min \{ 0.21,\,\,   \max  (    0.09 - 0.27 \log M,\,\,
0.037 + 0.033 \log M ) \},
\end{equation}
where $ c = -0.055(\log M-1.3)^{2} $ for $ \log M >1.3 $,
otherwise $ c = 0 $. When a star evolves along the MS,
its outer layers tend to expand as the core contracts.
The value of $ k $ can be described as a function of the current radius
$ R $ and the radius $ R_{0} $ at ZAMS,
\begin{equation}
k \simeq (k_{0} - 0.025) \left(\frac{R}{R_{0}}\right)^\alpha +
0.025\left(\frac{R}{R_{0}}\right)^{-0.1},
\end{equation}
where
\begin{equation}
  \alpha  = \left\{
    \begin{array}{ccc}
      -2.5  & \mbox{} & \log M < 0,  \\
      -2.5 + 5 \log M & \mbox{for} & 0 < \log M <0.2,   \\
      -1.5  & \mbox{ } & 0.2 < \log M.
    \end{array}  \right.
\end{equation}
The internal rotational profile of a star is set to be rigid rotation in the BSE code,
same as in \citet{d13}.

\textit{Tidal interaction.} Tidal torques in binary stars tend to
synchronize the stellar rotation with the orbital motion. The
efficiency is critically dependent on the ratio of the stellar
radius to the binary separation \citep{z77,h81}, and the effect of
tides is strong in short-period systems.
If the synchronization time scale
is less than the MS lifetime of the Be star,
the Be star may synchronize its rotation with
the orbital motion,  losing its Be character
before leaving the MS \citep{rl98}.
This will reduce the formation rate of
Be stars in narrow systems.

\textit{Mass transfer.} We follow the treatment of
\citet{d13} on the transfer of mass and angular momentum. During
RLOF the transferred material may either form an accretion disk
around the secondary or directly impact on its surface, depending on
the minimum distance $ R_{\rm min} $ compared with
the accretor's radius $R_2$ \citep{ls75}.
If the $ R_{\rm min} <R_2$, the stream
impacts directly on its surface, and the specific angular momentum
of the impact stream is $\sim (1.7GM_{2}R_{\rm min})^{1/2}$.
Otherwise the mass flow misses the accretor and collides with itself
at a larger radius, after which the viscous process leads to the
formation of a Keplerian accretion disk. The inner edge of the
accretion disk will stretch inward until contact with the surface of
the star, and the specific angular momentum of the transferred
matter can be expressed as $ (GM_{2}R_{2})^{1/2}$.

There are two types of mass transfer processes associated with the
Be star formation: the mass transfer without the CE occurrence and
the post-CE mass transfer\footnote{The pre-CE mass transfer and CE
evolution usually proceed so rapidly that the secondary hardly
accretes any matter.}. The former (termed as channel 1) has be
discussed by some authors \citep{p91,pz95,bv97}. When the primary
overflows its RL, if the mass transfer proceeds stably and a CE
stage is avoided, transfer of matter will be able to spin up the
secondary to be a Be star; the remnant of the primary will evolve to
be a helium burning star (He star), and finally a compact object
(WD, NS or BH). The latter (termed as channel 2) involves the CE
evolution. If the system survives the spiral-in phase, the left
binary will contain a He star and a MS star. The He star then
evolves and fills its RL once more, and transfers material to the
secondary, leading to the formation of a Be star \citep{b09}.

\textit{Mergers of two MS stars.} Unstable mass transfer will cause
the binary to enter a CE stage or come into contact and coalescence.
For CE evolution, if the orbital energy is not enough to unbind the
primary's envelope, the secondary will merge into the primary. If
the two MS stars become contact, the binary is also assumed to merge
into a MS star\footnote{ Stars evolved off the MS stage have
developed a core in its center, and the product of the merged binary
will remain this core so that it does not belong to a Be star.}. We
follow \citet{h02} to account for the products of mixing and
mergers. After the merger, the product will settle into thermal
equilibrium on a thermal timescale and we assume that it can
efficiently lose the excess angular momentum, forming a fast
rotating star.
However, the original prescription of \citet{h02} assumes that
the product of mergers involving two MS stars is completely
mixed and no mass is lost during this process. This may
underestimate the convective core mass for massive post-merger
MS stars. Here we assume that a certain fraction
$ \mu_{\rm loss} $ of the total mass of the binary system
is lost during the merger. In our calculation,
$ \mu_{\rm loss} = 10\%$ is adopted, in line
with \citet{l95} and \citet{d13}.

\section{Results}
The evolution of a primordial binary is determined by four initial
parameters: the primary mass $M_{1}$, the secondary mass $ M_{2} $,
the separation $a$ (or orbital period $P_{\rm orb}$), and the
eccentricity $ e $. The initial  eccentricity has minor effect on
the population synthesis results \citep{h02}, thus we assume
circular orbits here for simplicity. In our calculations, the masses
of the primary are chosen to be in the range of $ 1.5 M_{\odot}$ to
$60 M_{\odot}$, since the remnants of stars with $M_{1}\geqslant
60M_{\odot} $ are extremely rare according to the initial mass
function (IMF). The secondary masses are set to be between $0.1
M_{\odot} $ and $25 M_{\odot} $ to ensure that almost all of the Be
stars are contained, with a flat distribution of $1/q$. For the
initial orbital separation, we assume that $\ln a$  is evenly
distributed between $a=3R_{\odot}$ and $10^{4}R_{\odot}$. We adopt
solar metallicity Z = 0.02 and an IMF with a
power-law exponent of $-2.7$ \citep{k93}. A constant star formation
rate ($ S = 5 M_{\odot}\, {\rm yr}^{-1} $) is assumed over the past
15 Gyr.

In Table 1 we present the calculated numbers of binaries containing
a Be star with a He star/WD/NS/BH companion ($ N_{\rm BeHe} $, $
N_{\rm BeWD} $, $ N_{\rm BeNS}$, and $ N_{\rm BeBH} $,
respectively), and of isolated Be stars originating from disrupted
Be/NS and Be/BH systems, and from mergers of two MS stars ($ N_{\rm
dBeNS} $, $ N_{\rm dBeBH}$ and $ N_{\rm merger}$, respectively) in
the Galaxy. We see that the numbers of Be binaries drop
significantly with the increasing mass of the compact stars. The
value of $ N_{\rm BeHe} $ varies drastically from $\sim 10^{5} $
(channel 1) to $ < 100 $ (channel 2), suggesting that very few Be/He
systems can be produced after the CE evolution. The number from
mergers of two MS stars is zero in channel 2, because this situation
does not happen at all. Detailed results on the formation of Be
stars are described below.

\subsection{Be/NS binaries}

The formation  of Be/NS binaries has been
investigated by many authors
\citep[e.g.,][]{rh82,hr87,p91,pz95,bv97,r01,b09}.
There are 81 confirmed BeXRBs in our Galaxy, and 48 of them host a NS
\citep{l06,reig11}. For most of them the spectral types
of Be stars and the orbital periods are
known \citep{n98,m08}, so we can compare the calculated
results with the observations, and present possible constraints
on the formation processes of BeXRBs.

Figure~2 shows the distribution of the masses ($ M_{\rm Be} $) of Be
stars  in Be/NS binaries in the blue solid lines. The grey solid
lines represent the distribution derived from observations
\citep[data are taken from][]{reig11}. The left, middle and right
panels correspond to the results with Models I, II and III,
respectively. For each model, the top and bottom panels reflect the
results from channels 1 and 2, respectively. The blue dashed lines
denote the accretion fraction $ f $, i.e., the ratio of the average
accreted masses over stars in each bin and the Be star mass.

We first discuss the distribution of $ M_{\rm Be} $ in the
top panels obtained from channel 1. The Be stars hardly accrete any
more material after reaching the break-up limit in Model I, so $ f
\lesssim 0.1$, and a large fraction of the Be stars tend to have
relatively low mass. In Model II, the Be stars can accrete half of
the transferred mass from the donor, so $ M_{\rm Be} \gtrsim 8
M_{\odot} $  and $ f\sim 0.3-0.5 $. In Model III, more masse can be
accreted, thus $ M_{\rm Be} \gtrsim 13 M_{\odot} $ and $ f
\gtrsim 0.45 $. From Model I to Model III, the parameter space for stable
mass transfer becomes smaller, the fraction of accreted material
becomes larger, hence the predicted numbers of Be/NS binaries reduce
from $ \sim 1800 $ to $ \sim 100$, and the minimal masses of the Be
stars increase from around $3 M_{\odot}$ to around $13 M_{\odot} $.
The distributions of $ M_{\rm Be}$ in the bottom panels (from
channel 2) are roughly similar, ranging from around $3 M_{\odot}$ to
around $10-13 M_{\odot}$. The main reason is that, after CE
evolution, the remaining He star is generally less massive than the
accretor, so the transferred matter is relatively small, and $ f <
0.2$ always holds for the three models. Obviously the predicted
distribution of Be/NS binaries through channel 1 of Model II seems
to best fit the observations.

Figure~3 shows the  $ P_{\rm orb} $ distributions in the three
models for systems formed through channels 1 and 2,
respectively\footnote{In Fig.~3, $ P_{\rm orb} $ is cut off at the
maximal orbital period  of 1000 days, because Be/NS systems with
longer periods may not be observed as XRBs due to the very low
accretion luminosity..
The orbital periods of Be/NS binaries in channel 1 are mainly
distributed around
 $10 - 1000$ days (the solid lines), while those in channel 2 have relatively
shorter periods, peaked around a few days to tens of days (the
dashed lines), because of orbital shrink during the CE phase}. The
observational orbital periods of the BeXRBs in the Galaxy lie in the
range of $ \sim 10-300 $ days \citep[e.g.,][and references
therein]{b09,c14}, more compatible with the results from channel 1.


It is well known that the spectral types of isolated Be stars in our Galaxy
can be late than A0 (corresponding to $\sim 3\,M_{\sun}$), but
in BeXRBs the Be stars are more massive than $ \sim 8 M_{\odot}$
\citep{n98,m08}.
In order to explain this difference,
\citet{p95} assumed that evolution of a close binary with initial
mass ratio larger than 2.5 would not produce any Be stars, because
they do not transfer any mass but rather evolve towards a CE phase.
\citet{pz95} instead suggested mass loss from the $ L_{2} $ point in
binary systems. The basic idea is that the related effective angular
momentum loss can promote the binary to evolve into a CE stage, and
only binaries with  $M_2> 8 M_{\odot}$ can survive. In our approach,
we have taken into account both the mass transfer stability and the
possible effect of mass loss under different conditions. We find
that only in channel 1 of Model II, the calculated mass distribution
of Be stars is in line with observations, but for Be/NS binaries
formed from channel 2 of the same model, the distributions of both $
M_{\rm Be} $ and $ P_{\rm orb} $ disagree with the
observation\footnote{We need to caution that the observed limit of
about $8\,M_{\odot}$ is affected by the rather uncertain mass
determination of Be stars, as well as by possible incompleteness of
the observed sample of BeXRBs.}. We will address this issue below.

\citet{h10} found that the lowest velocity that a star has to reach
to show the Be-effect may vary as a function of the stellar mass.
Low-mass ($ < 4 M_{\odot} $) B stars need to rotate extremely fast
($ V_{\rm eq}/V_{\rm cr}\gtrsim 0.96$, where $V_{\rm eq}$ is the
rotational velocity at the equatorial plane) to create an outflowing
disk, while for massive stars ($ > 8.6 M_{\odot}$), the threshold
drops to $ V_{\rm eq}/V_{\rm cr} \sim 0.63  $.
In Fig.~4 we plot the number distributions of Be/NS systems in Model
II as a function of $ M_{Be} $, with the black and red curves
corresponding to the threshold value of $V_{\rm eq}/V_{\rm cr}= 0.8$
and 0.95, respectively. We find that, with a higher $V_{\rm
eq}/V_{\rm cr}= 0.95$, the number of low- and intermediate-mass Be
stars can be reduced slightly, but there are still too many such
Be/NS binaries. The reason is that a star can be spun up to critical
rotation by accreting about $(10-15)\%$ of its original mass
\citep{p81}, and this can be practically satisfied by most post-CE
mass transfer.

Here we propose two possible ways to remove intermediate-mass Be
stars in Be/NS systems. (1) To form a Be star, there should be
enough mass accreted by the secondary star with $ f > 0.2 $. When
calculating the stellar rotation, we assumed that the star is
rigidly rotating, while differential rotation may be more realistic.
In addition, when the accretor spins up to close to the critical
limit, it may lose mass due to the effect of the centrifugal force
\citep{p05}. Thus more accreted mass is needed to turn a B type star
into a Be star. The accreted mass during the post-CE mass transfer
process is relatively small with $ f $ always $ < 0.2 $. If this
threshold works, then low- and intermediate-mass Be stars would not
be produced in this channel. (2) The effect of magnetic braking
combining strong magnetic field (as in Bp stars) and the enhanced
wind could spin down the stars. This model was established by
\citet{d10} and \citet{dsd13} for the spin angular momentum
evolution of the accretors in Algol-type binary stars. A
differentially rotating star might generate magnetic fields in its
radiative atmospheres. The processes of accretion may also induce
strong winds which interacts with the magnetic field ($ \gtrsim  $ 1
KG). Due to magnetic braking, the rotational velocity of the
accretor may be reduced to below its critical limit and lose the
character of a Be star. The problem with this explanation is that no
magnetic field has been reliably detected in any Be star
\citep{wg12}.

Based on the above arguments we do not favor the formation of Be/NS
binaries through the post-CE mass transfer. In the following we only
discuss the Be stars formed through channel 1. As the mass
distribution of Be stars in Be/NS binaries can fit the observation
better in Model II, we regard it as the standard model and discuss
the results in this model in the following, if not mentioned
otherwise. However, we note that the assumption of 50\% accretion
efficiency in Model II is completely ad hoc, and Models I and III
are actually more physical, considering the roles of rotation and
thermal equilibrium of the accretor to constrain the accretion
processes. The fact that Model II can better reproduce the BeXRB
population indicates that some important physics is still lacking in
the treatment of the binary evolution. Meanwhile,  the obtained
results also depend on the adopted ways of mass loss. All the three
models assume isotropic re-emission of the material that is not
accreted.

Figure~5 displays the expected distribution of Be/NS binaries in the
$ P_{\rm orb} - e $ plane. In the left and right panels we plot the
binaries with the NSs originating from electron-capture SNe and
core-collapse SNe, respectively.  During core-collapse SNe the
newborn NSs experience violent explosions and a high-velocity kick,
the resulting binaries tend to have eccentricities $>0.5$. In the
case of electron-capture SNe, the binary eccentricities $\lesssim
0.7 $. The eccentricity distribution may provide important
information about the two subpopulations of BeXRBs resulting from
different types of SNe \citep[][see however, Cheng et al.
2014]{k11}.

\subsection{Be/BH binaries}

There is currently only one confirmed Be/BH binary in the Galaxy,
i.e., MWC 656  \citep{cn14}, which contains a BH of mass $ 3.8-6.9
M_{\odot} $ in a $ \sim 60$ day orbit. From our BPS calculations,
the number of Be/BH systems is $ \sim 10 $ in the standard model,
but increases to $ \sim 250 $  in Model I. In Model III, no Be/BH
can be formed, because in the case of stable mass transfer with $
M_{1} > 20 M_{\odot}$, the secondary mass has been increased to $ >
22 M_{\odot} $. Figure 6 shows the distribution of Be/BH binaries in
the $ P_{\rm orb}- M_{\rm Be}$ plane, with the left and right panels
corresponding to Models I and II, respectively (the dashed line
denotes the orbital period of NWC 656). The masses of the Be stars
can be as low as $ \sim 8 M_{\odot} $ in Model I, while in Model II
$M_{\rm Be}  \gtrsim 18 M_{\odot}  $. \citet{b09} have explored the
formation of the populations of Be/NS and Be/BH binaries, and found
the numbers of these two types of systems to be $579 - 1578$ and $19
- 82$, respectively. The numbers of Be/BH systems from their
calculation are covered by ours. In particular, the ratio of Be
binaries with NSs to the ones with BHs is $\sim 7-53$ (in Models I
and II), while the preferred value is $\sim 30-50$ in \citet{b09}.
The difference results from different definitions of BeXRBs and
different treatments on the formation of Be stars. For example,
\citet{b09} assumed that a constant fraction of B type stars are Be
stars, and that all binaries either survive CE evolution or evolves
to a merger if the donor star is in the HG.

Finally we emphasize that in both \citet{b09} and this work
semi-analytic formulae \citep[e.g.][]{f12} are adopted to estimate
the BH masses, which are assumed to be largely determined by the
progenitor mass\footnote{Also note that in Eq.~(16) in \citet{f12},
$a_1$ should be $0.25-1.275/(M-M_{\rm proto})$ rather
$0.25-(1.275/M-M_{\rm proto})$.}. However, it has been shown that
successful SN explosions are intertwined with failures in a complex
pattern that is not well described by the progenitor initial mass
and is not simply related to compactness \citep[][and references
therein]{k14,pt14}. For example, progenitors with certain initial
masses less than $20\,M_{\sun}$ are likely to form BHs rather NSs.
This may remarkably influence the predicted number of BHs and the BH
mass functions.

\subsection{Be/He and Be/WD binaries}

Most of the Be binaries contain a He star or a WD companion (see
Table 1). In Figs.~7 and 8 we plot the distributions of Be/He and
Be/WD binaries in the $ P_{\rm orb}-M_{\rm Be} $ plane,
respectively. The mass distributions of Be stars in the Be/He and
Be/WD binaries are shown in Fig.~10 with red and green curves,
respectively.

Our calculations show that a Be star can have a He companion
with mass $ \sim 0.3-17 M_{\odot}$. When
the He star's mass is less than
$ \sim 2.5 M_{\odot} $, the rejuvenated MS
lifetime of the Be star becomes
shorter than that of the He star, and such binaries contribute
most to the population of Be/He binaries.
During the formation of Be/compact star binaries,
the primaries also spend some time in the He star stage when
their envelopes are stripped, although these He stars are
more massive and evolve quickly into
compact stars before the Be stars evolve off
the MS. The orbital periods of Be/He binaries lie
between $\gtrsim 10$ days and $\sim 200$ days, and cluster
around $ \sim20 $ days.
Based on the combined energy distribution
of a B2V star and a $ 1 M_{\odot} $ He star companion,
\citet{p91} showed that the luminosity of the binary in  XUV
should be dominated
by the He star. At lower frequencies the contribution
of the He star is negligible and it is difficult to detect
these systems.

The orbital periods of Be/WD binaries also range from $\sim 10$ days
to $\sim 200$ days, and peak around $\sim20$ days. They tend to
possess intermediate-mass Be stars, similar as Be/He binaries.
\citet{r01} found that the peak of the $P_{\rm orb}$ distribution is
$ \sim 100 $ days and there are very few systems with $ P_{\rm
orb}\lesssim  30$ days. The differences mainly result from the
synchronization mechanisms adopted. In \citet{r01}, relatively
short-period Be/WD binaries are removed from the population due to
the operation of the synchronization mechanism of \citet{t87}. This
mechanism is more efficient than that suggested by \citet{z77}
adopted in the BSE code, so stars with $ P_{\rm orb}\sim 10- 30$
days can be synchronized and lose the Be character.


Our calculations suggest that there may be $\sim 10^5$ Be/WD
binaries in the Galaxy.
It is interesting to note that currently there
are no Be/WD binaries observed in the Galaxy, though three were
identified in the Large and Small Magellanic Clouds \citep{k06,s12,l12}.
The circumstellar disk of the Be star
may be truncated by the tidal torque from the WD,
because of its circular orbit \citep{al94,no01,z04},
with little accretion onto the WD. In this case the WD's
UV and optical emission powered by cooling
could be detected for hottest WDs. If the truncation is inefficient
and the WD can accrete
matter from the Be star disk, it might experience episodes of shell burning
(as in nova systems), appearing as a transient supersoft X-ray source.
However, its XUV and soft X-ray radiation
is likely to be  absorbed by the gas in the envelope of the Be star in which
the WD is embedded  \citep{a91}. Analyses by \citet{n13} and
\citet{wp13} suggest that small amount of circumstellar matter
local to the WD can easily suppress its X-ray emission.

\subsection{Isolated Be stars}

The masses of isolated Be stars can be as low as $ 3 M_{\odot} $
\citep{s82}. Here we consider their formation through binary
evolution in two ways. The first is from ``disrupted Be/NS and Be/BH
systems'' because of the SN explosions. When the primary star
evolves to experience a SN explosion and leaves a NS or BH, the
binary system may be disrupted and produce an isolated Be star. Be
stars originating from disrupted Be/NS systems are much more
numerous than from Be/BH systems (see Table 1), because of the IMF
and larger amplitude kicks imparted on the NSs. The mass
distributions of the calculated (in the standard model) and observed
isolated Be stars are plotted in Fig.~9 in the blue and grey solid
lines, respectively. The observational data were taken from
\citet{s82} with a magnitude limit of $ V \leq 6 $, so many
late-type Be stars might have been missed due to the selection
effect, and the actual distribution of isolated Be stars may have a
even lower mass peak. The number of isolated Be stars originating
from disrupted Be/NS systems in the standard model is estimated to
be $ \sim 5200 $, about 10 times the number of the surviving Be/NS
systems. The masses of the Be stars are generally larger than $ \sim
8 M_{\odot}$, similar as in Be/NS systems.

The second and much more important formation channel of isolated
Be stars is the merger of two MS stars.
In the
standard model, the number of Be stars formed through mergers
is $ \sim 1.9\times10^{6}$, similar as in Models I and III.
This is much more than from the disrupted Be/NS systems. More
importantly, the predicted Be stars tend to have low masses,
compatible with the observation (we need to mention that in the
observed sample in Fig.~9 (also in Figs. 10 and 11 below) is
obviously incomplete, so we can only compare the shapes of the
distributions), suggesting that mergers are more promising in
forming isolated Be stars.

\section{Discussion and conclusions}

In this paper, we investigate the formation of Be stars
through mass transfer and mergers in binaries.
In Fig.~10 we present the mass distributions of all Be stars in
binaries,
and of isolated Be stars formed from disrupted Be/NS and Be/BH
systems. The thick blue line reflects the overall distribution of
these Be stars. For $M_{\rm Be}<10M_{\sun}$, it is dominated by Be
stars with a He star and a WD companion. More massive Be stars are
likely to be isolated Be stars from disrupted Be/NS systems.

How important is binary interaction in the formation of Be stars?
Since mergers of two MS stars may also produce Be stars, we compare
in Fig.~11  the mass distributions of Be stars formed through
channel 1 and mergers in the standard model, as well as all B type
stars with the blue, green, and black lines, respectively.
Their numbers are correspondingly $ \sim 5.7\times10^{5} $, $ \sim
1.9\times10^{6} $, and $\sim 1.17\times10^{7} $. Obviously these
numbers depend on the fraction of massive binaries, IMF, and the
initial mass ratio distribution. The fraction of binaries in the
Galaxy has been shown to be $ 69\%(\pm9\%) $ with orbital periods
between $ 10^{0.15}$ days and $ 10^{3.5} $ days \citep{sd12,d13}. So
we take the binary fraction to be between 50\% and 100\%, and plot
the calculated fraction of Be stars in B type stars in Fig.~12.
\citet{p91} concluded that no more than 60\% of the population of Be
stars are formed through case B binary evolution. \citet{bv97} found
that a minority of the Be stars (less than 20\% and possibly as low
as 5\%) are due to close binary interaction. Our results show that,
combining the effects of both mass transfer and merger, the fraction
of Be stars in B type stars can reach  $ \sim13\%-30\% $, compatible
with the observational results of $ 20\%-30\% $ \citep{z97} and $
1/5-1/3 $ \citep{mg05}. We emphasize that all the numbers of Be
stars cited in this work should be taken as upper limits, because
not all rapidly rotating B type stars exhibit the Be phenomenon.

 Finally we summarize our results as follows.

(1) By considering different possible mass accretion
histories for the mass gainer in a binary, we calculate the critical mass ratios
for stable mass transfer. We find that in Be/NS binaries the Be star
masses and orbital periods are consistent with observations if
they are formed by stable and nonconservative mass transfer (i.e.,
channel 1 of Model II).

(2) There are about $10^6$ isolated Be stars in the Galaxy
originating from both disrupted Be/NS systems and mergers of two MS
stars, but the latter play a much more important role.

(3) The ratio of Be/NS binaries to the ones with BHs can be as small
as $\sim 7$, suggesting that there could exist a hidden population
of Be/BH binaries in the Galaxy.

(4) Both Be/He and Be/WD binaries tend to have low-mass Be stars
and orbital periods of tens of days.
Most of the He stars in Be/He binaries  are less massive
than $ \sim 2.5 M_{\odot}$.

(5) The fraction of Be stars resulting from binary evolution among B
type stars is around $13\%- 30\%$.

\acknowledgements
We thank the referee for constructive comments which
have greatly helped improve the manuscript.
This work was supported by the Natural Science Foundation of China
under grant numbers 11133001, 11203009 and 11333004, the Strategic
Priority Research Program of CAS (under grant number XDB09000000), and the
graduate innovative project of Jiangsu Province (CXZZ13-0043).


\appendix

\section{Two Cases of Evolutionary Sequences}

To illustrate how the stability of mass transfer depends on the
evolutionary state of the binary system, in Fig.~A1 we show the
evolutionary tracks of a binary in Model I with the initial
parameters of $ M_{1}=10 M_{\odot} $ and $ q =3$. Prior to the mass
transfer, the primary has lost some of its mass in a stellar wind,
leading to a slight widening of the orbit. In the top panel, the
initial orbital period $ P_{\rm orb} $ is set to be 3 days. At an
age $ \sim $ 22.16 Myr, the primary, still on the MS, overfills its
RL and commences mass transfer. The orbital period decreases from
3.2 day to 2 days after $\sim 1 M_{\odot}$ mass has been transferred
to the secondary. Meanwhile, the mass transfer rate rises
 to $ \sim 10^{-3} M_{\odot}\, {\rm yr}^{-1}$.
At this time the radius of the secondary exceeds its RL radius and
the binary becomes contact. In the middle panel, the initial $P_{\rm
orb} $ is taken to be 15 days. The primary has evolved off the MS
and entered the shell-burning phase when RLOF initiates. The mass
transfer proceeds rapidly but stably  at a rate $ \sim 10^{-4}-
10^{-2}M_{\odot}\, {\rm yr}^{-1}$. A small amount ($\sim\,0.4
M_{\odot}$) of the transferred material is able to spin up the
secondary into a Be star, and the rest of the material is assumed to
be ejected out of the system. The orbital period decreases until the
primary's mass drops to $ \sim 5 M_{\odot}$. After that the mass
ratio reverses, and $P_{\rm orb}$ increases to $ \sim 13 $ days at
the end of the evolution. After the envelope of the primary is
stripped, a $ \sim 2 M_{\odot} $ He core is left. In the bottom
panel, the initial $ P_{\rm orb} $ is 320 days. When the primary
overfills its RL, it has climbed to the red giant branch. Once RLOF
starts, the mass transfer rate rises to $ \sim 10^{-2} M_{\odot}\,
{\rm yr}^{-1}$ within $\lesssim 100$ yr. The mass transfer proceeds
on the dynamical timescale and a subsequent spiral-in stage is
followed.

In Fig.~A2 we present similar evolutionary sequences for a binary in
Model II with $ M_{1}=10 M_{\odot} $ and $ q =2$. The initial $
P_{\rm orb}$ are 2, 3, and 5 days in the top, middle and bottom
panels, respectively. In the top panel, at the onset of RLOF (at an
age of $ \sim $ 18.86 Myr), the primary is in the MS stage. When the
orbital period decreases to less than $ \sim1.6 $ days and the mass
transfer rate increases to $\gtrsim 10^{-4}M_{\odot}\, {\rm yr}^{-1}
$, the secondary fills its RL, leading to a contact phase. In the
middle panel, the mass exchange begins at an age $ \sim $ 21.86 Myr
when the primary is also a MS star. The binary experiences stable
mass transfer at a rate $ \sim 10^{-5}-10^{-3}M_{\odot}\, {\rm
yr}^{-1} $, with a temporary phase of detachment. Half of the
transferred material is ejected from the binary,  so the secondary
accretes $ \sim 4 M_{\odot}$ mass. The resulting binary consists of
a 2 $ M_{\odot} $ He star and a $  \sim 9 $ $ M_{\odot}$ Be star. In
the bottom panel, RLOF occurs (at an age $ \sim $ 23.53 Myr) when
the primary has evolved to the HG. The mass transfer rate increases
to $ \sim 10^{-3} M_{\odot}\, {\rm yr}^{-1}$ and the orbit shrinks
to less than 4 days. The secondary quickly fills its RL followed by
a CE phase.

\clearpage

\begin{figure}

\includegraphics[scale=0.4]{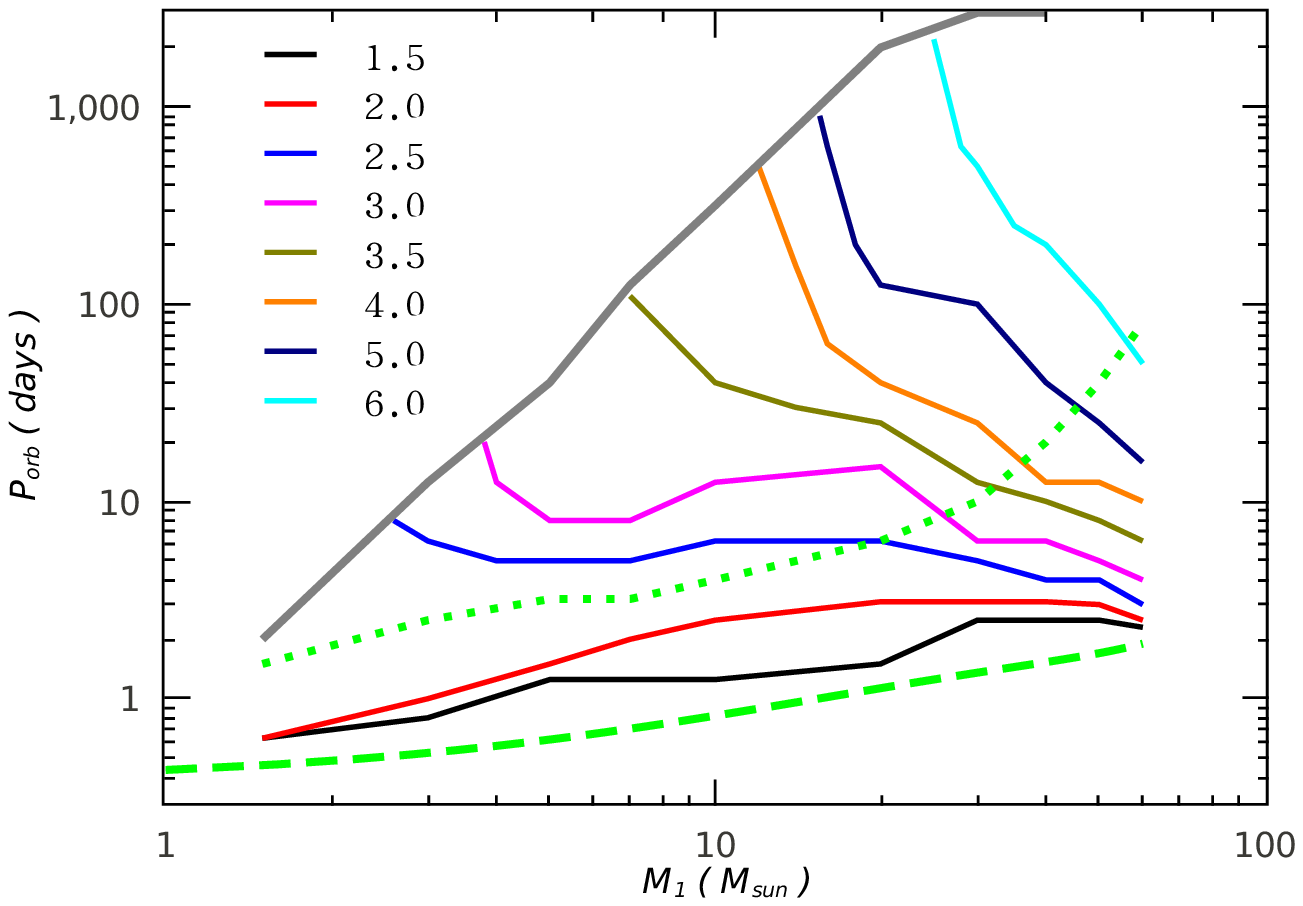}\includegraphics[scale=0.4]{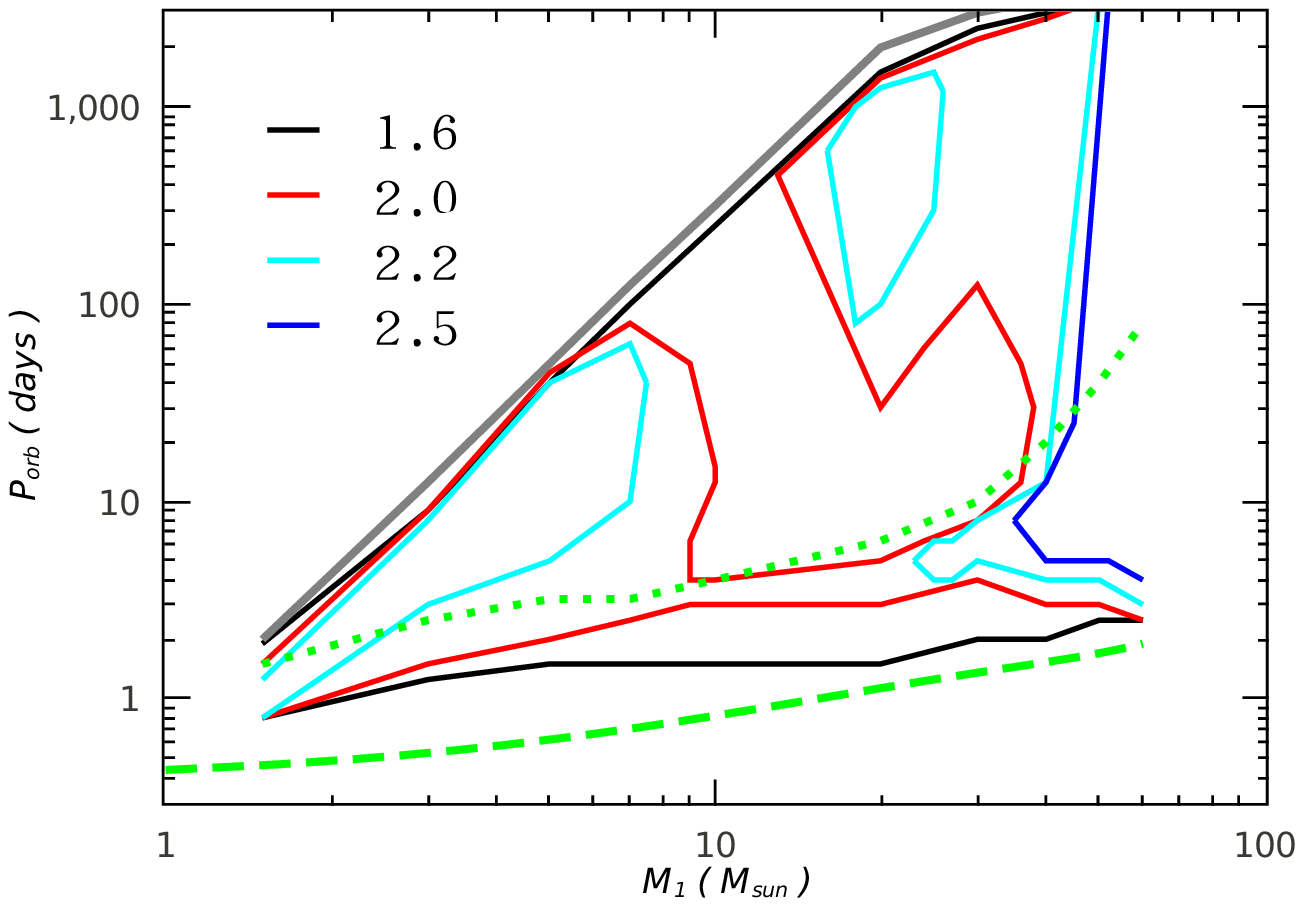}
\includegraphics[scale=0.4]{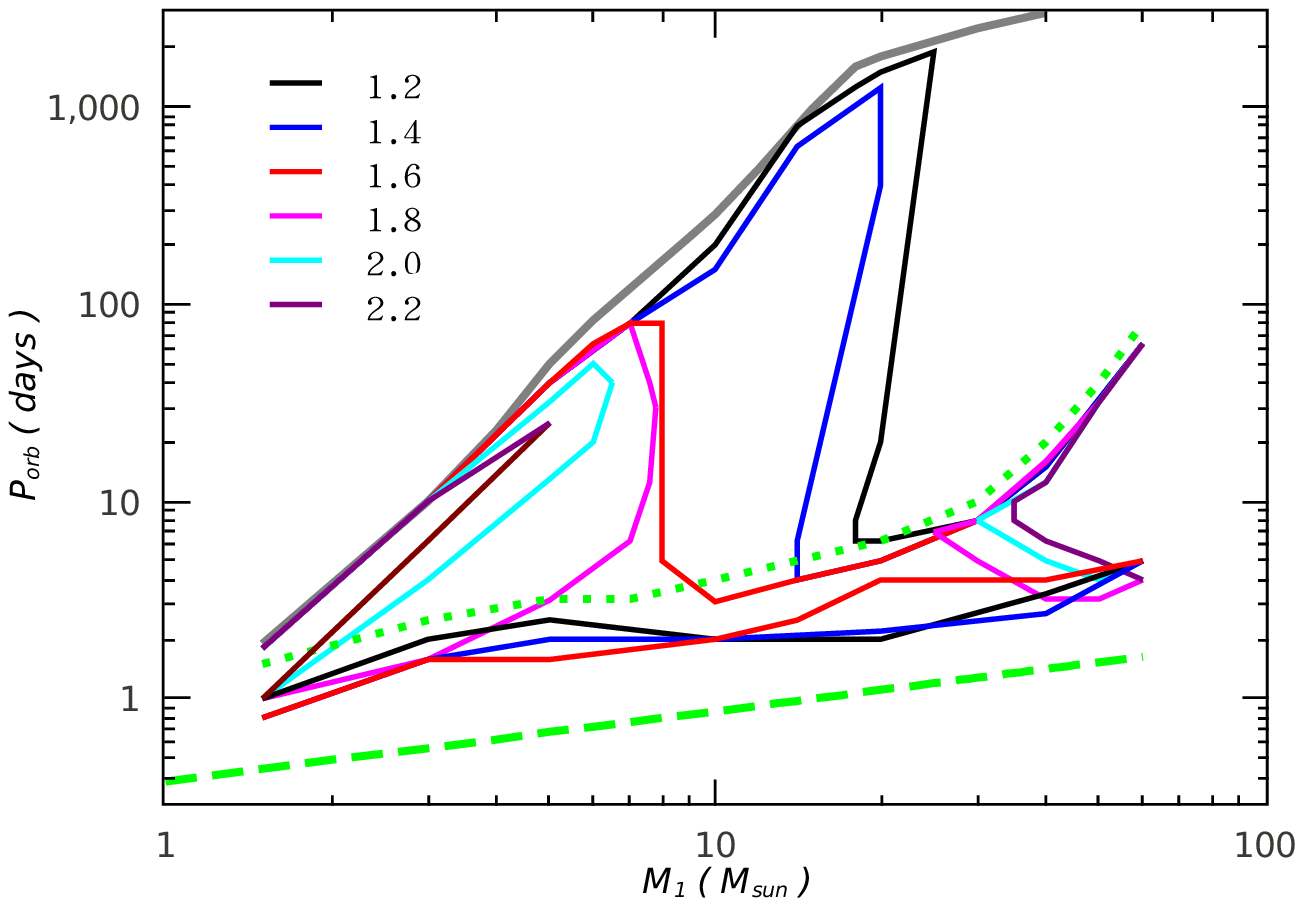}
\caption{The solid curves describe the allowed parameter space in
the initial $ P_{\rm orb}-M_{1} $ plane for stable mass transfer, in
which contact and CE phases can be avoided. The left, middle and
right panels correspond to Models I, II and III, respectively. In
each panel, the number next to each colored label denotes the
initial mass ratio ($q=M_1/M_2$) of the binary components. The green
dashed, green dotted and black solid curves represent the orbital
periods when the primary overflows its RL at ZAMS, the end of MS,
and the end of HG, respectively (they depend very weakly on $ q $,
and here we adopt $q =2$).       \label{figure1}}

\end{figure}




\clearpage

\begin{figure}

\centerline{\includegraphics[scale=0.6]{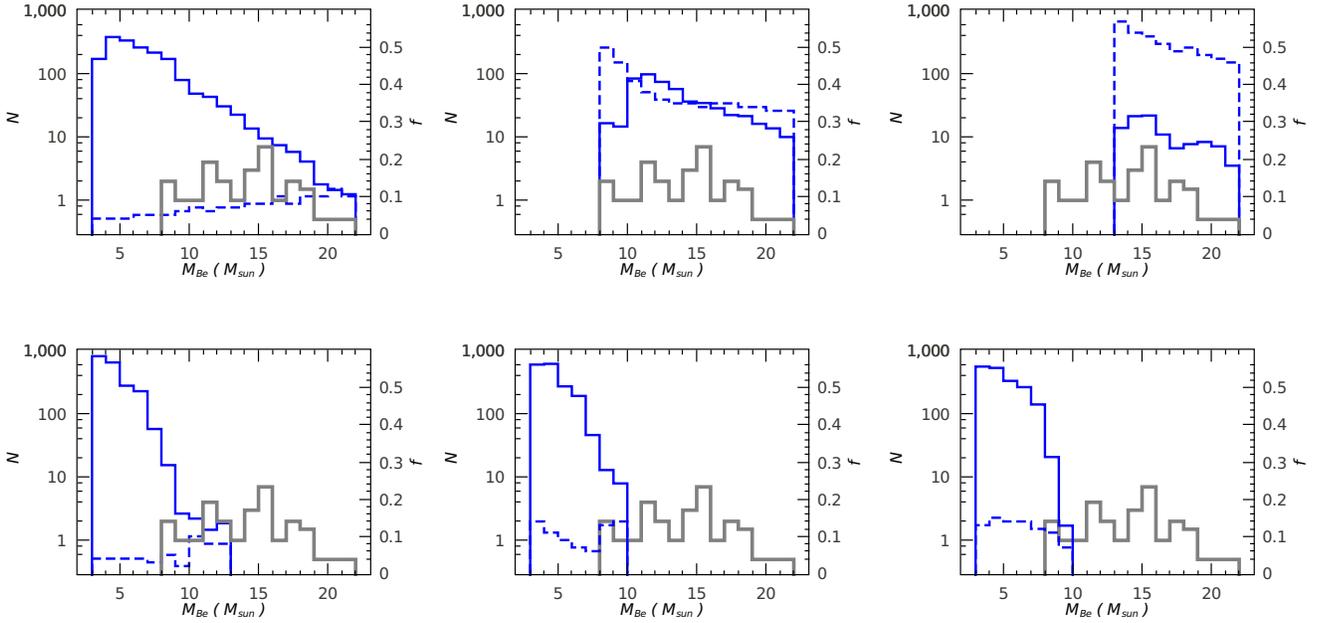}}
\caption{The blue solid lines represent the mass distributions of Be stars in Be/NS binaries
formed through channels 1 (top panel) and 2
(bottom panel). The left, middle and right panels correspond to
Models I, II and III, respectively. The blue dashed lines reflect
the fraction $ f $ of the average accreted mass by the
Be stars in each bin. The grey dashed lines show the
observational distribution of Be stars with a NS companion. Here the
spectral type data of Be stars are taken from \citet{reig11}. We
calibrate the relation between the spectral types and the masses from \citet{hh81}.
 \label{figure2}}

\end{figure}

\clearpage

\begin{figure}

\includegraphics[scale=0.4]{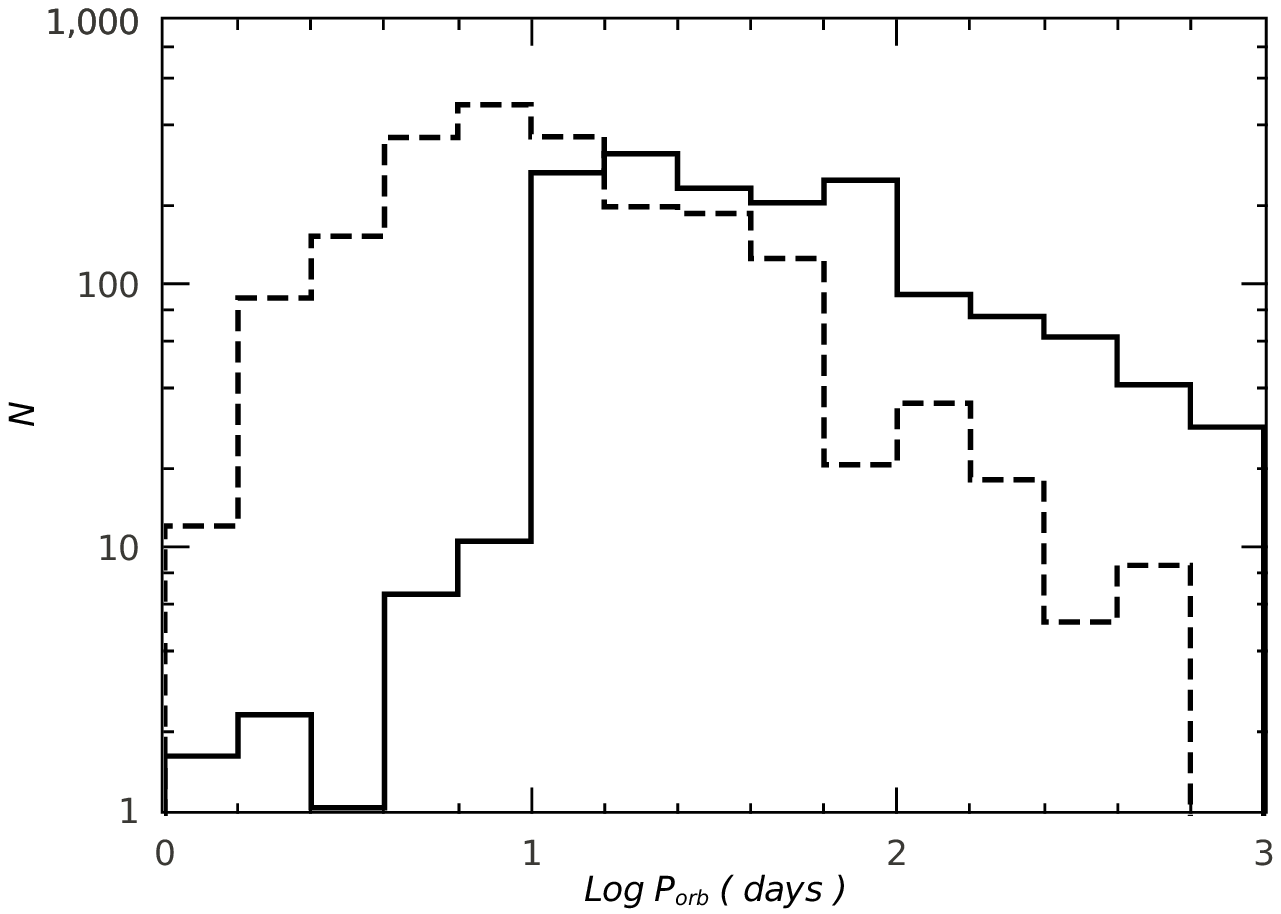}
\includegraphics[scale=0.4]{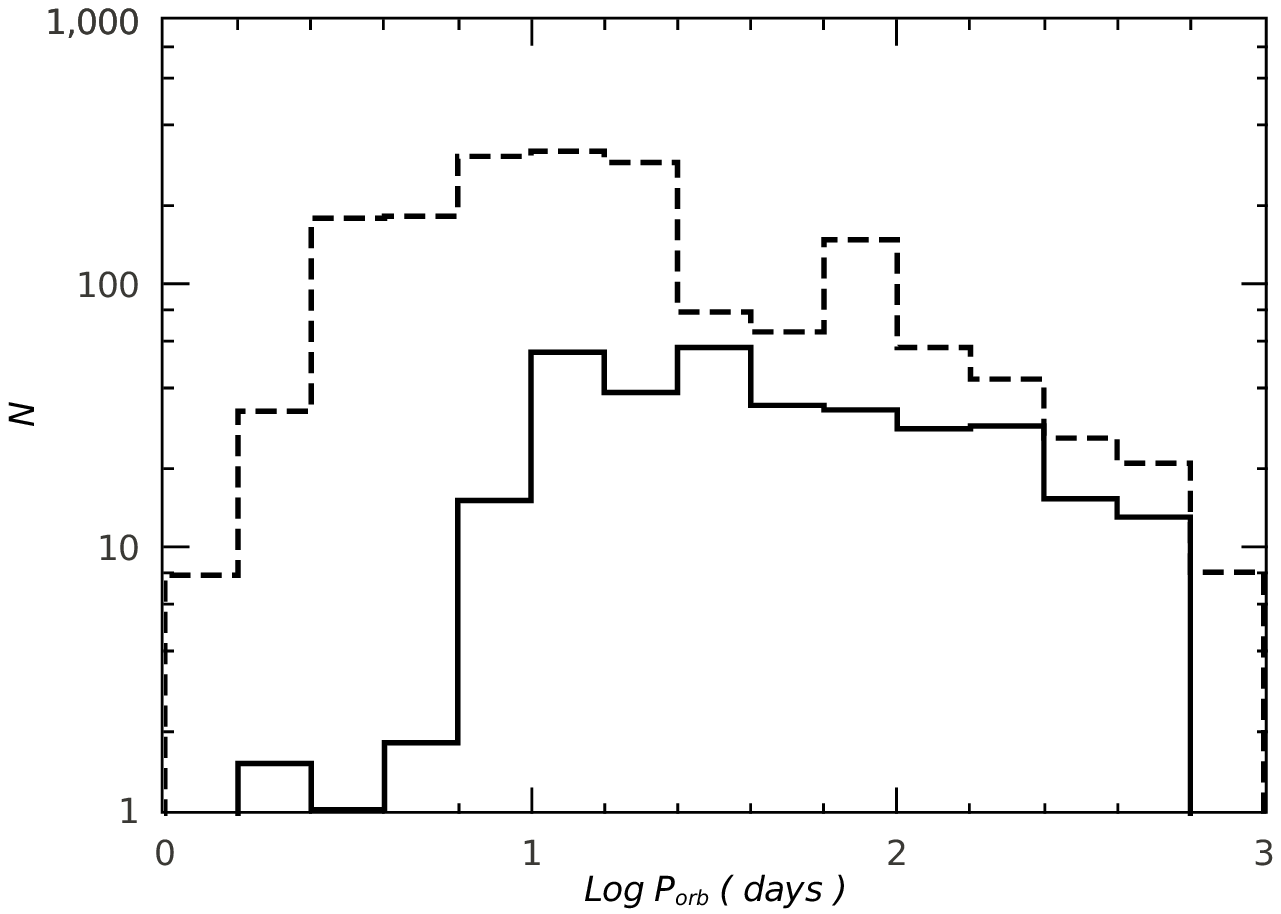}
\includegraphics[scale=0.4]{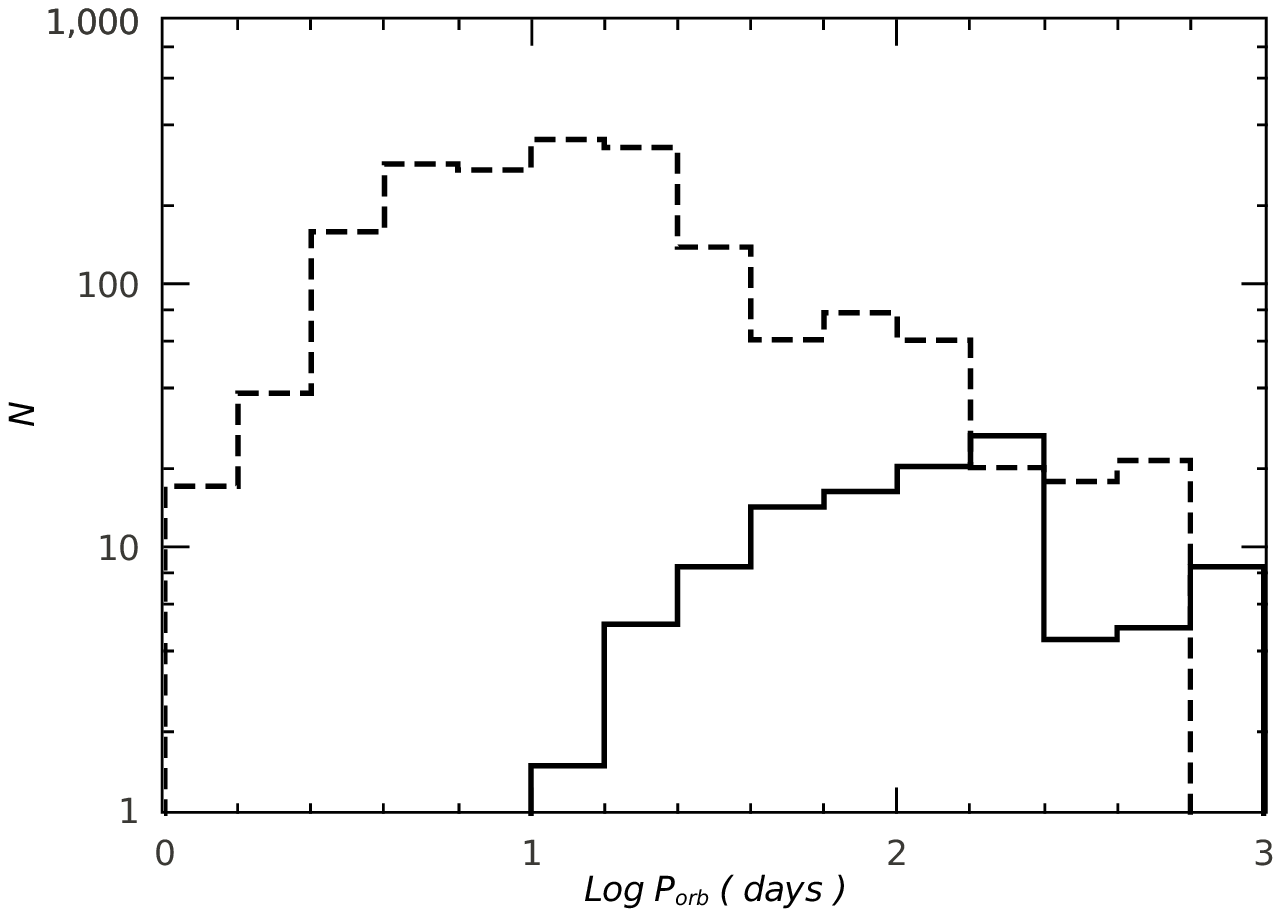}
\caption{The distributions of the orbital periods of Be/NS binaries.
The solid and dashed curves correspond to the systems formed through
channels 1 and 2, respectively. The left, middle and right panels
correspond to Models I, II and III, respectively.
 \label{figure3}}

\end{figure}

\clearpage

\begin{figure}

\includegraphics[scale=0.8]{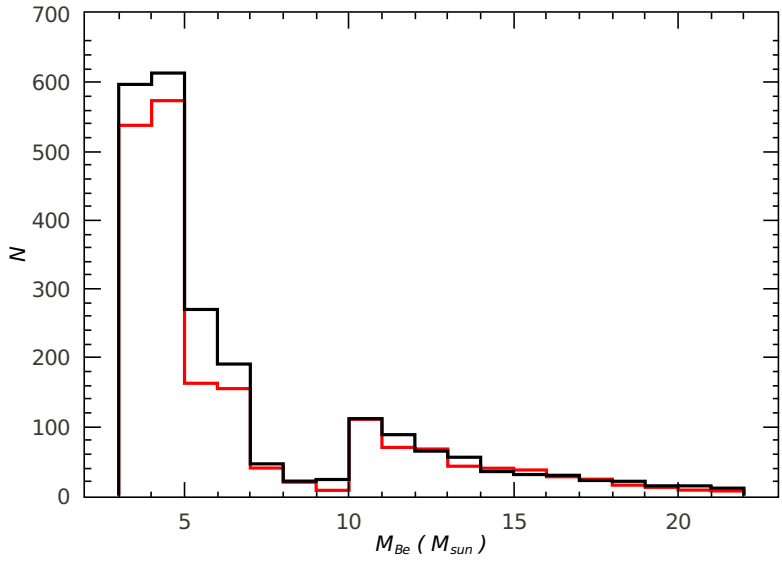}
\caption{The distributions of the masses of Be stars in Be/NS
systems in the standard model. The black and red curves correspond
to the parameter $ V_{\rm eq}/V_{\rm crit} = 0.8$ and 0.95,
respectively .
   \label{figure5}}

\end{figure}

\clearpage

\begin{figure}

\includegraphics[scale=0.6]{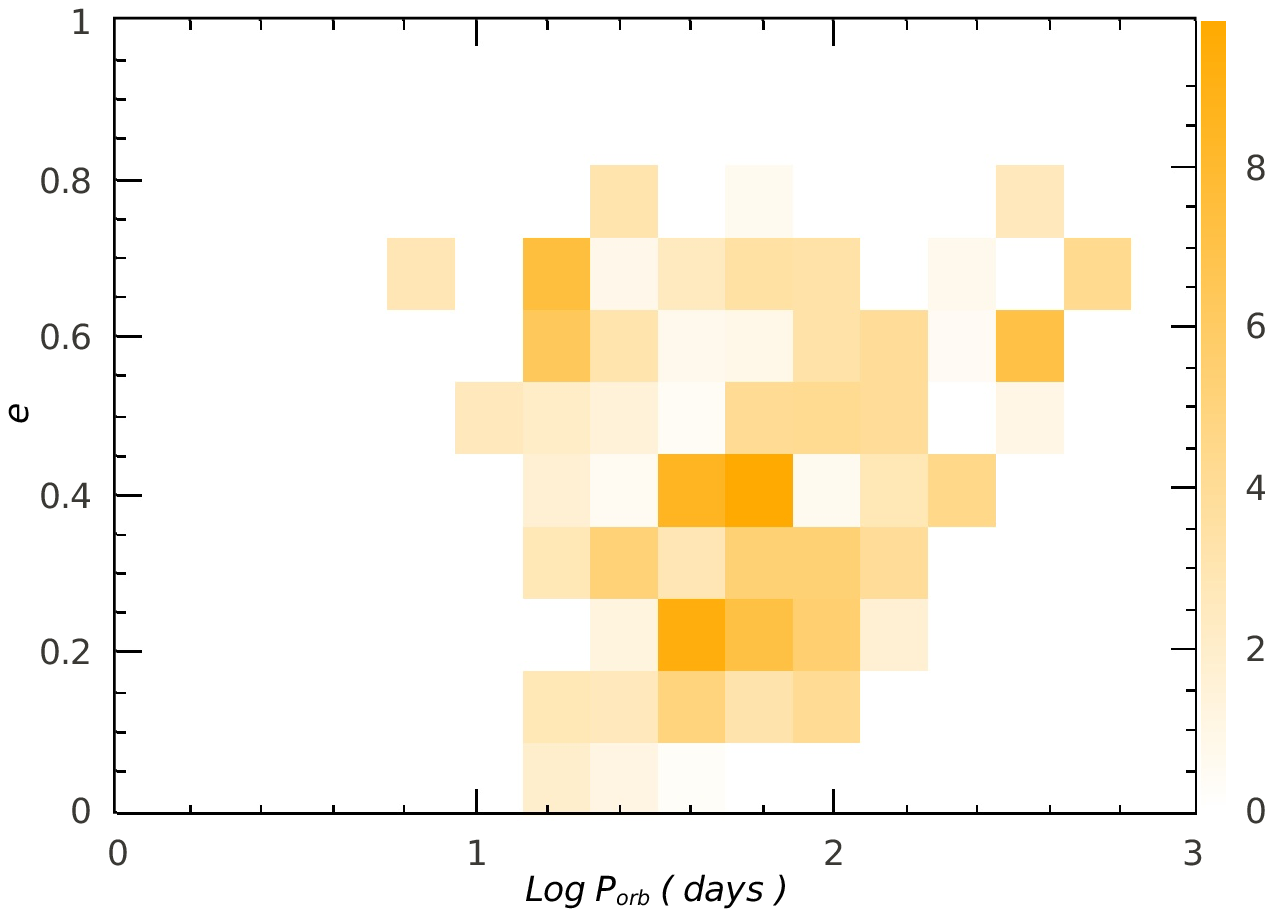}
\includegraphics[scale=0.6]{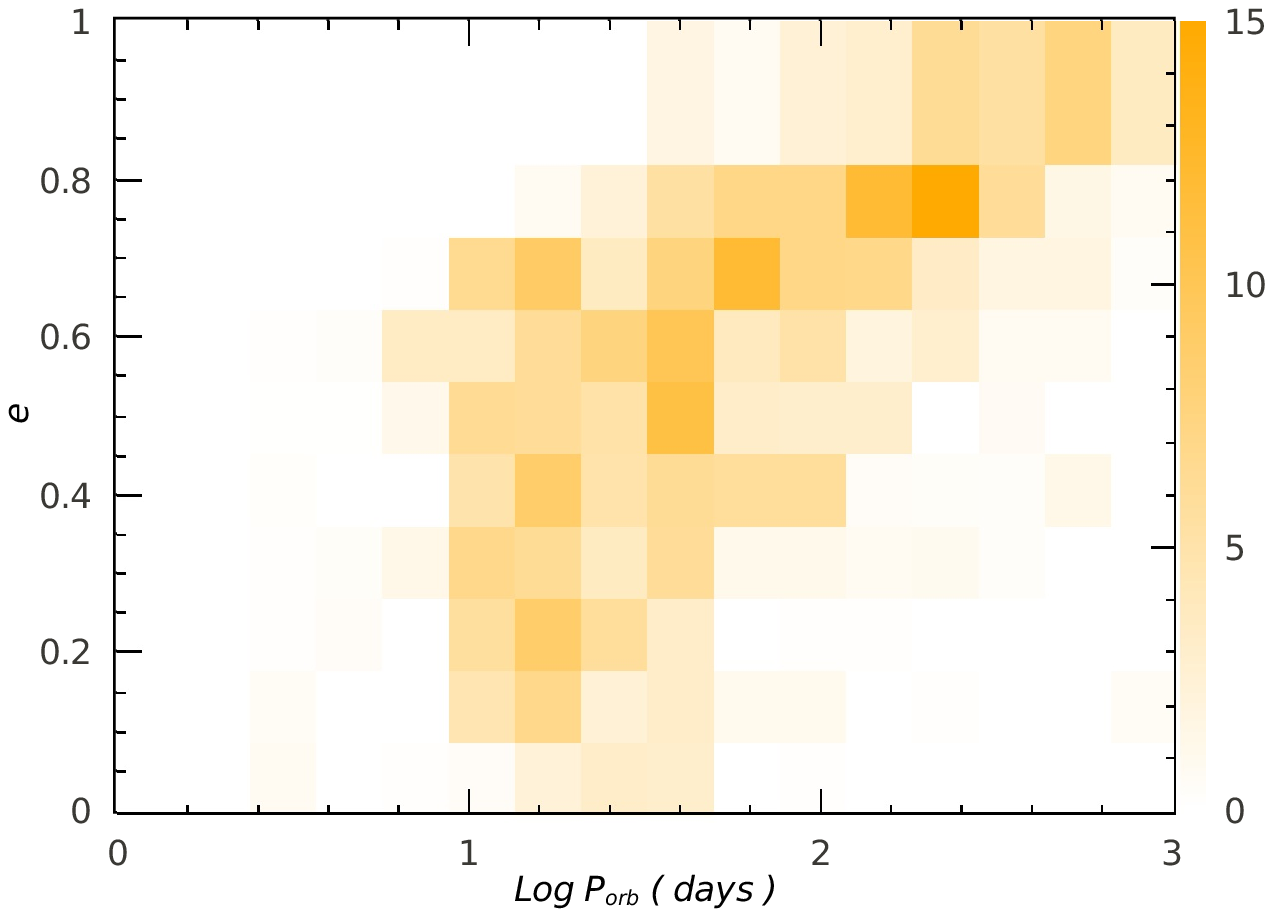}
\caption{The distribution of Be/NS binaries in the
$ P_{\rm orb}- e $ plane, formed in the standard model.
In the left and right panels the NSs  originate from
electron-capture SNe and core-collapse SNe, respectively.
\label{figure6}}

\end{figure}

\clearpage

\begin{figure}

\includegraphics[scale=0.6]{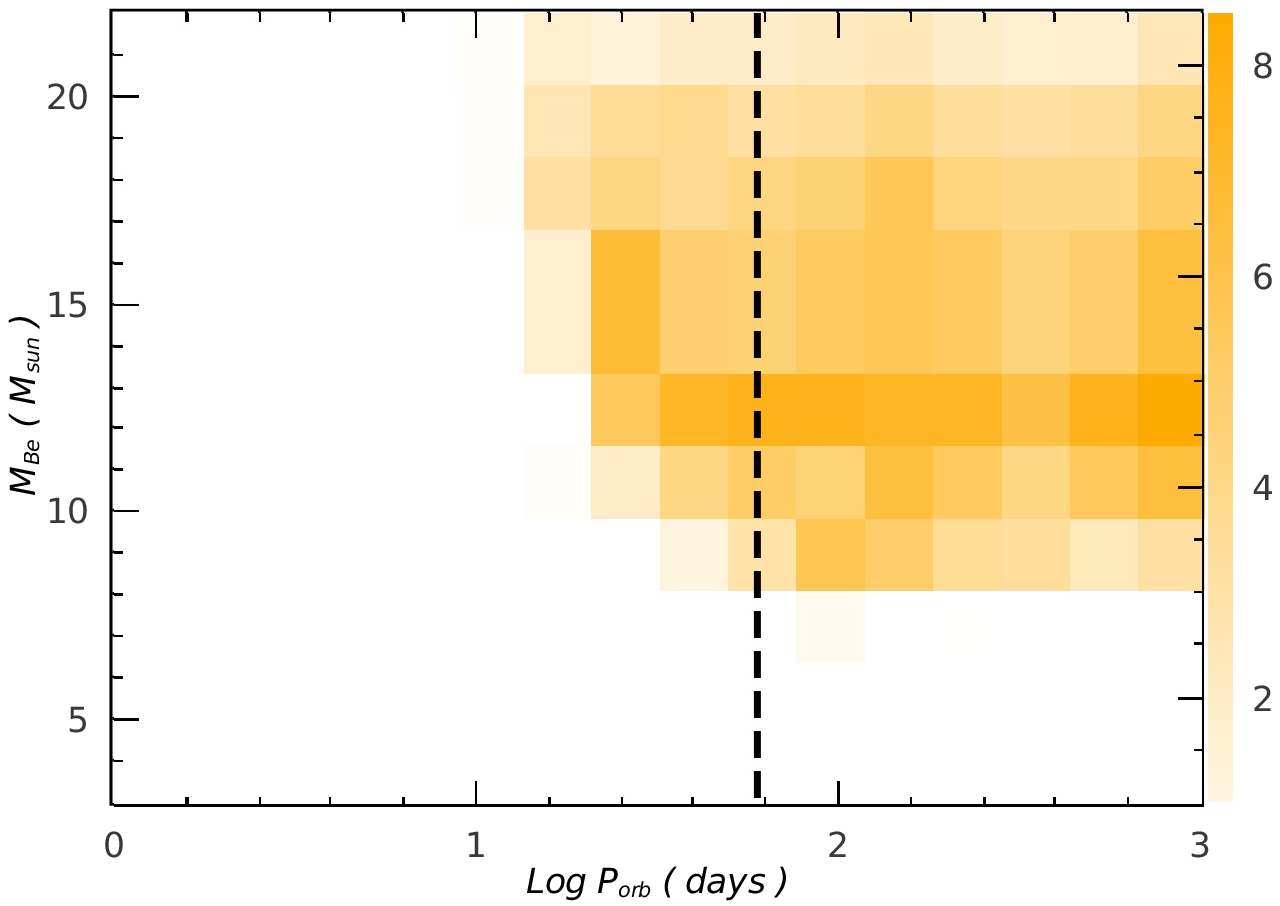}
\includegraphics[scale=0.6]{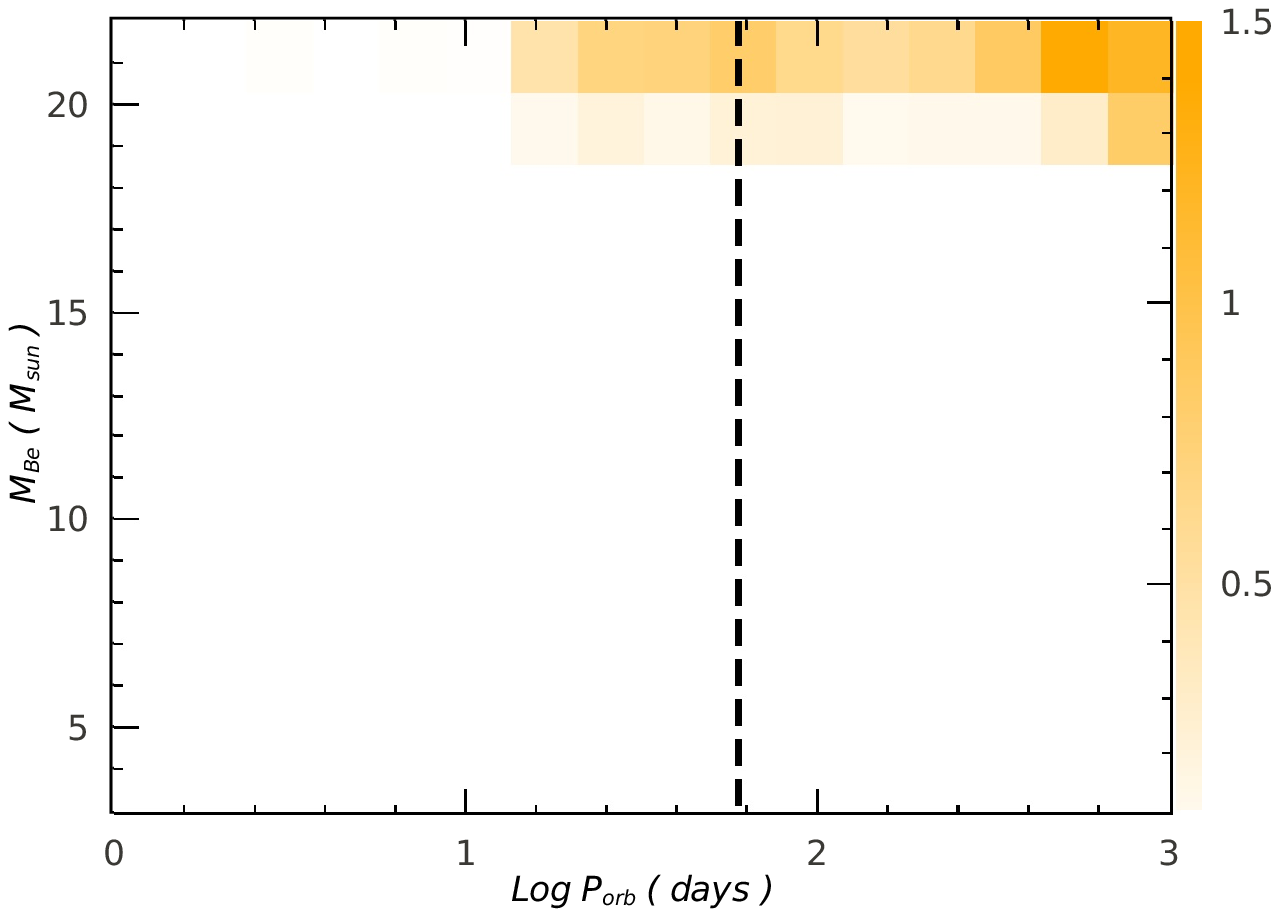}

\caption{The distribution of Be/BH binaries  in the
$ P_{\rm orb}-M_{Be} $ plane in Models I (left panel) and II (right panel).
The dashed line shows the orbital period of the Be/BH binary MWC 656.
\label{figure6}}

\end{figure}

\clearpage

\begin{figure}

\includegraphics[scale=0.8]{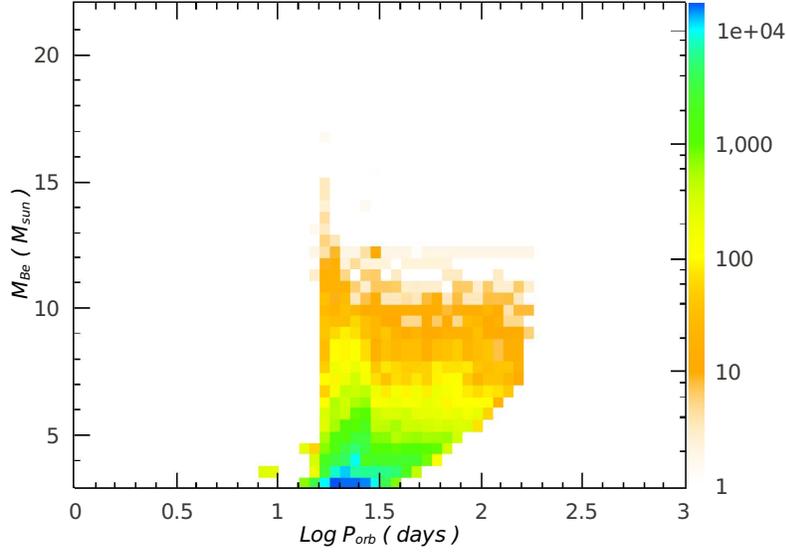}
\caption{The distribution of Be/He binaries  in the
$ P_{\rm orb}-M_{Be} $ plane in the standard model.
   \label{figure7}}

\end{figure}

\begin{figure}

\includegraphics[scale=0.8]{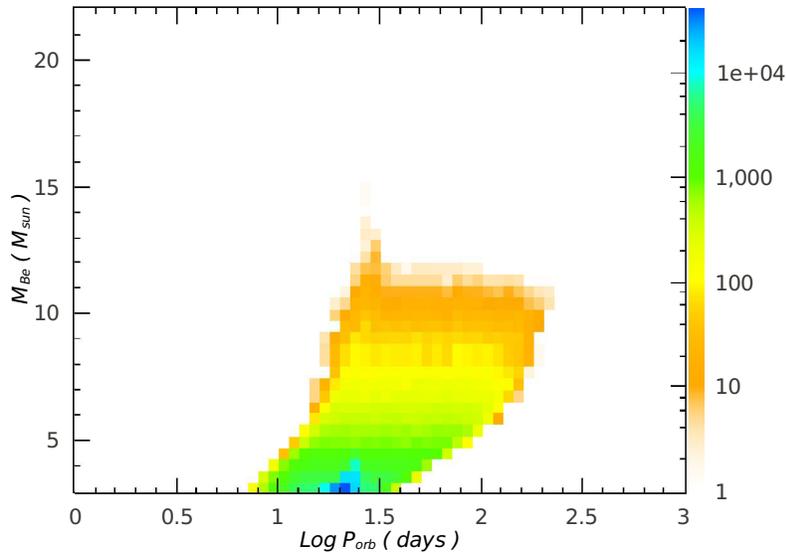}
\caption{Same as Fig.~7, but for  Be/WD binaries.
 \label{figure8}}

\end{figure}
\clearpage

\begin{figure}

\includegraphics[scale=0.8]{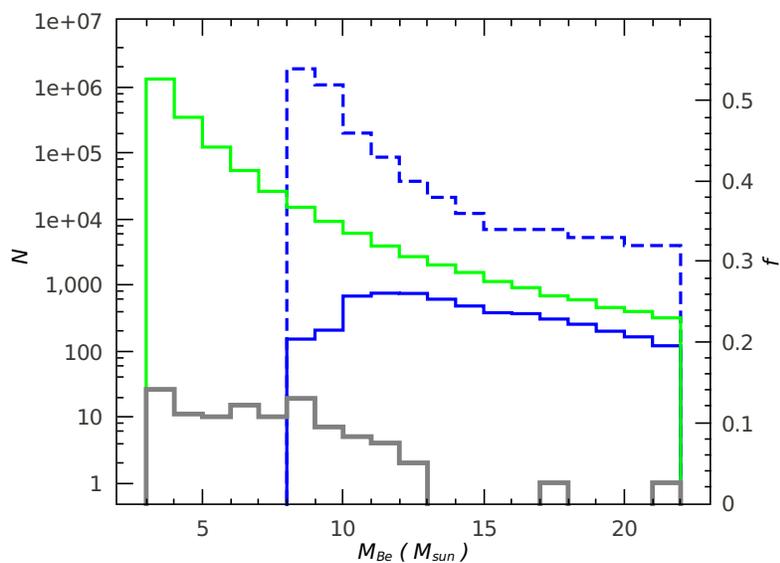}
\caption{The blue and green solid lines represent the mass
distribution of isolated Be stars evolved from disrupted Be/NS
binaries and mergers of two MS stars in the standard model,
respectively. The blue dashed line denotes the fraction $ f $ of
average accreted mass by the Be stars in each bin in the first case.
Observational distribution of isolated Be stars is shown in the grey
line, with the spectral type data of Be stars taken from \cite{s82}.
   \label{figure9}}

\end{figure}

\clearpage
\begin{figure}

\includegraphics[scale=0.8]{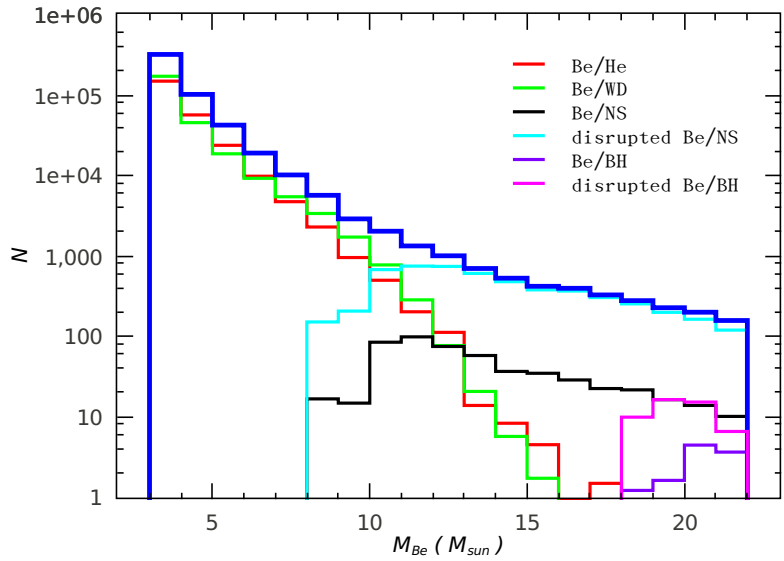}
\caption{The mass distribution of all Be stars evolved from the
standard model, shown in the blue line. Other \textbf{six} lines
represent the distributions of Be stars in Be/He, Be/WD, Be/NS, and
Be/BH binaries, and isolated Be stars from disrupted Be/NS and Be/BH
systems.
   \label{figure10}}

\end{figure}

\clearpage

\begin{figure}

\includegraphics[scale=0.8]{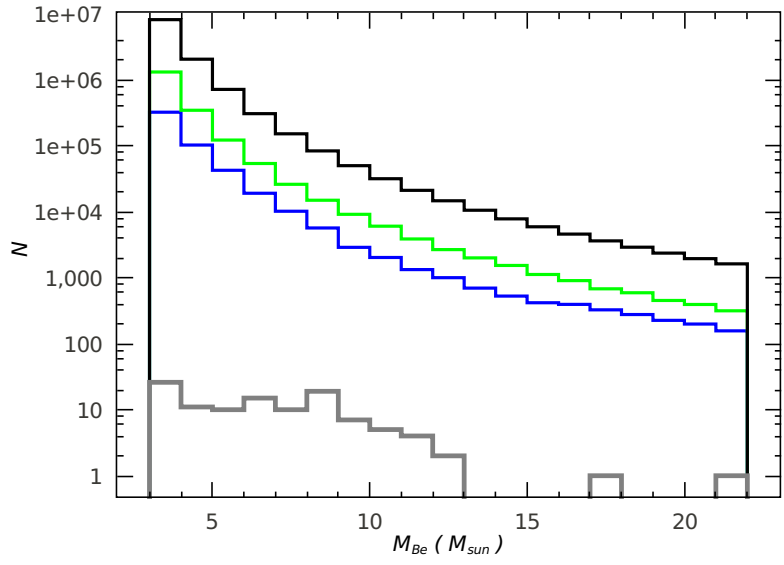}
\caption{The black, blue, and green lines represent the mass
distributions of all B type stars, Be stars formed from channel 1
and from MS mergers in the standard model, respectively. The grey
line shows the observational distribution of isolated Be stars in
the Galaxy. \label{figure11}}

\end{figure}

\clearpage

\begin{figure}

\includegraphics[scale=0.8]{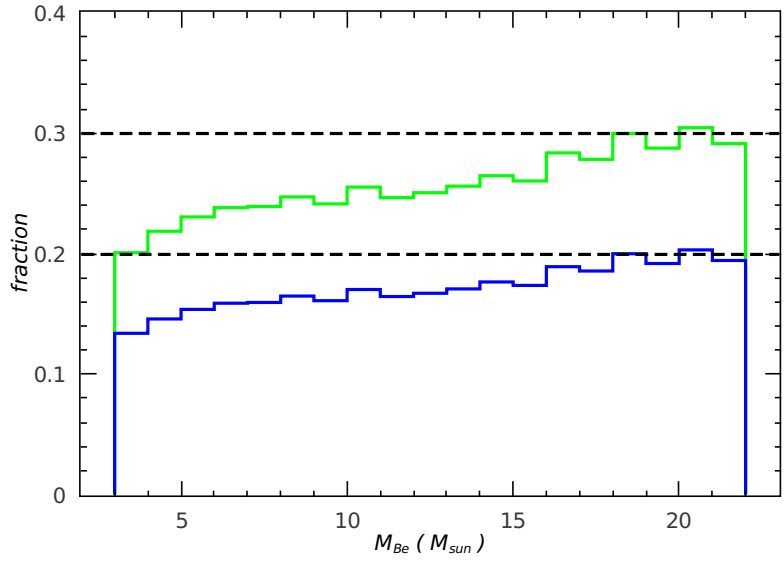}
\caption{The fraction of Be stars in B type stars. The
green and blue solid lines are obtained under the assumption that the
fraction of primordial binaries
in the Galaxy are 100\% and 50\%, respectively. The two dashed lines
show the range of derived fractions from observations \citep{z97}.
\label{figure12}}

\end{figure}
\clearpage

\begin{figure}
\figurenum{A1}
\centerline{\includegraphics[angle=-90,width=1.0\textwidth]{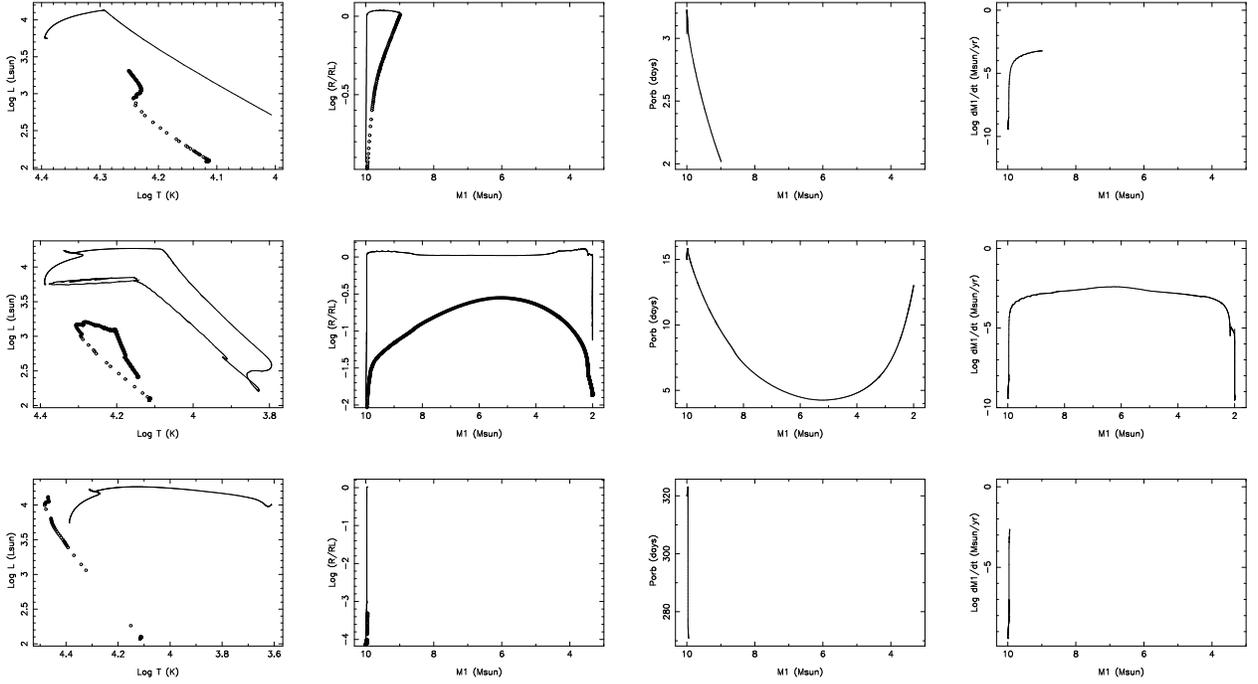}}
\caption{Evolution of the binary systems with $ M_{1} = 10 M_{\odot}
$, $ q = 3 $, and $ P_{\rm orb} $ = 3 day (top), 15 days (middle),
and 320 days (bottom) in Model I. Panels from left to right are
shown the evolutionary tracks of the primary \textbf{(solid curve)
and the secondary (dotted curve) }in the H-R diagram, the ratios of
the primary (solid curve) and secondary (dotted curve) radii to
their RL radii, the orbital period, and the mass transfer rate.
\label{figure14}}

\end{figure}

\begin{figure}
\figurenum{A2}
\centerline{\includegraphics[angle=-90,width=1.0\textwidth]{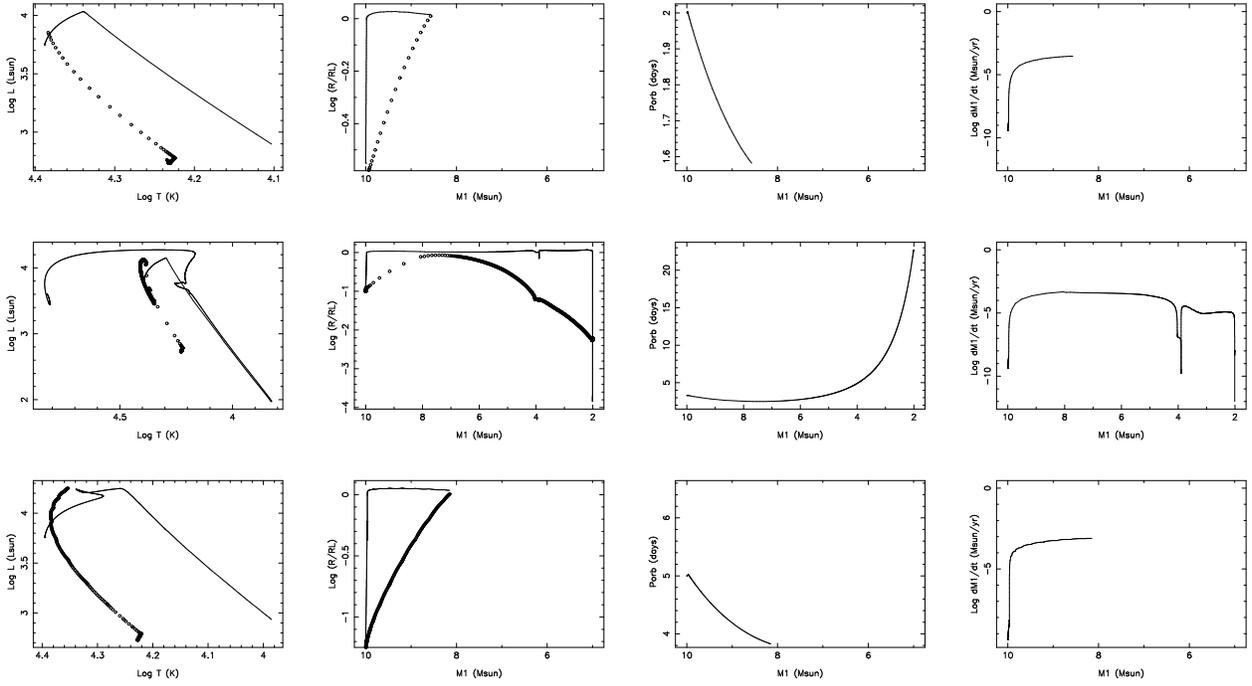}}
\caption{Same as Fig.~A1, but for binary systems with $ q = 2 $ and
 $ P_{\rm orb}  = 2$ days (top), 3 days (middle), and 5 days
(bottom) in Model II.  \label{figure15}}

\end{figure}

\clearpage

\begin{table}
\begin{center}
\caption{The predicted numbers of Be/He, Be/WD, Be/NS, and Be/BH binaries and
 isolated Be stars from disrupted Be/NS, Be/BH binaries, and
 MS star mergers in
 the Galaxy in the three models and the two different channels.\label{tbl-1}}
\begin{tabular}{lcccccccc}
\\
\hline
 Models &  I & I& II &II& III&III   \\
\hline
 Channels & 1 &2 & 1 & 2 & 1 & 2   \\
\hline
$ N_{\rm BeHe} $ & $6.5\times10^{4}$& $ <  100$& $2.5\times10^{5}$& $ <  100$& $4.1\times10^{5}$& $ <  100$ \\
$ N_{\rm BeWD} $ & $1.8\times10^{5}$& $1.3\times10^{5}$& $2.6\times10^{5}$& $1.4\times10^{5}$& $1.3\times10^{5}$& $1.6\times10^{5}$ \\
$ N_{\rm BeNS} $ & 1800 & 2043 & 531 & 1761& 102 & 1842 \\
$ N_{\rm dBeNS} $ & $2.0\times10^{4}$ & 3930& 5246 & 2450& 1964 & 2505 \\
$ N_{\rm BeBH} $ & 245 & $ < $ 1 &   10 & $ < $ 1 & 0 & $ < $ 1 \\
$ N_{\rm dBeBH} $ & 573 & $ < $ 1 &   51 & $ < $ 1 & 0 & $ < $ 1 \\
$ N_{\rm merger} $ & $1.6\times10^{6}$ & 0 & $1.9\times10^{6}$ & 0& $2.1\times10^{6}$ & 0 \\

\hline
\end{tabular}
\end{center}
\end{table}

\end{document}